\documentclass[10pt]{elsart}

\usepackage[round,sort]{natbib}

\usepackage{graphicx}
\usepackage{epsfig}
\usepackage{color}

\usepackage{amssymb}
\usepackage{amsbsy}

\usepackage{lineno}

\usepackage{verbatim}

\def\os{\bar{\omega}_\sigma}
\def\ok{\bar{\omega}_\kappa}
\def\aniso{\mathcal{A}}
\def\dd{\mathrm{d}}
\def\pl{\partial}
\newcommand{\bs}[1]{\boldsymbol{#1}}
\def\thdn{\theta_i n_i}
\def\lan{\langle}
\def\ran{\rangle}
\def\al{\alpha}

\begin{document}

\begin{frontmatter}



\title{Design of three dimensional isotropic microstructures for maximized stiffness and
conductivity}


\author[UQ]{V. J. Challis\corauthref{vjc}},
\ead{vchallis@maths.uq.edu.au}
\author[UQ]{A. P. Roberts},
\ead{apr@maths.uq.edu.au}
\author[UQ,CSIRO]{A. H. Wilkins}
\ead{andrew.wilkins@csiro.au}

\corauth[vjc]{Corresponding author.  Telephone: +61 7 3346 1428.  Fax: +61 7
3365 1477}

\address[UQ]{Department of Mathematics, University of Queensland, Brisbane, QLD
4072, Australia}
\address[CSIRO]{Queensland Centre for Advanced Technologies, PO Box 883,
Kenmore, QLD 4069, Australia}

\begin{abstract}
The level--set method of topology optimization is used to design isotropic
two--phase periodic multifunctional composites in three dimensions.  One phase
is stiff and insulating whereas the other is conductive and mechanically
compliant.  The optimization objective is to maximize a linear combination of
the effective bulk modulus and conductivity of the composite.  Composites with
the Schwartz primitive and diamond minimal surfaces as the phase interface have
been shown to have maximal bulk modulus and conductivity.  Since these
composites are not elastically isotropic their stiffness under uniaxial loading
varies with the direction of the load.  An isotropic composite is presented with
similar conductivity which is at least 23\% stiffer under uniaxial loading than
the Schwartz structures when loaded uniaxially along their weakest direction.
Other new near--optimal isotropic composites are presented, proving the
capablities of the level--set method for microstructure design.  

\end{abstract}


\end{frontmatter}

\section{Introduction}
\label{intro}

The design of multifunctional materials is a growing field.
\citet{torquatoET02,torquatoET03} considered the simultaneous transport of heat
and electricity in three dimensional biphasic composites where one phase was
more thermally conducting and less electrically conducting than the other.  The
structures were required to be isotropic by imposing threefold reflection
symmetry.  When both thermal and electrical conductivity were maximized the
resultant composite resembled a Schwartz primitive (P) structure.\footnote{In
this paper Schwartz P and diamond (D) structures refer to composites with the
Schwartz P and D minimal surfaces as the phase interface.}  The optimality of
the Schwartz P structure and also of the Schwartz D structure for maximizing
both conductivity properties was shown numerically via finite element
calculations and using rigorous cross--property bounds.  Following this work,
\citet{torquatoET04} used finite element calculations to show the optimality of
Schwartz P and D structures for maximum bulk modulus and conductivity for any
ill--ordered phase properties (one phase has a larger conductivity but a smaller
bulk modulus and shear modulus than the other phase).  These publications
demonstrated that structures derived from Schwartz P and D minimal surfaces are
important multifunctional composites.  

\citet{guestET06} designed microstructures to maximize bulk modulus and fluid
permeability.  They demanded cubic elastic symmetry and isotropic flow symmetry
and used the solid isotropic material with penalization
\citep[SIMP,][]{bendsoe89,bendsoeET88} implementation of topology optimization.
Stokes' flow conditions were assumed to calculate fluid permeability and
optimized structures were presented for different coefficients in the
optimization objective.  For particular coefficients in the multiobjective
design problem they obtained structures very similar to the Schwartz P
structure, consistent with the results of \citet{torquatoET04}.  

The recent paper of \citet{kruijfET07}  considered optimal structures with
maximum stiffness and minimum resistance to heat dissipation and the design of
two dimensional composite materials with maximal effective thermal conductivity
and bulk modulus.  The two phases for the design problem were ill--ordered.  The
microstructures were required to be isotropic with respect to conductivity but
only square--symmetric with respect to elasticity.  The SIMP topology
optimization algorithm was used.  

In this paper, two-phase isotropic three dimensional periodic composites are
designed using topology optimization to have maximal bulk modulus and
conductivity.  The two isotropic phases for the microstructure design problem
are ill--ordered infinite--contrast materials: one of the phases has finite
stiffness but zero conductivity, whereas the other phase has finite conductivity
and zero stiffness.  The result is competition between the phases.  In
particular, for the composite to have nonzero stiffness and nonzero conductivity
both phases need to be connected.

Our study is motivated by a desire to design maximally stiff, electrically
conducting and isotropic cermets (composites of ceramic and metal), and
optimally stiff, porous and isotropic bone implants.  In the former situation,
the ceramic has high stiffness but low conductivity, whereas the metal has
comparatively low stiffness and high electrical conductivity.  For implants, the
implant material (titanium, for example) has high stiffness but is impenetrable
to bone in-growth, whereas the pore space has zero stiffness and allows bone
in-growth.  Here, the effective conductivity of the pore space is used to model
the ease of bone in-growth into the implant.  In reality, both scenarios will
also involve manufacturing constraints, we deal with the simplest microstructure
design problem and such constraints are left for future work.  The calculation
of electrical conductivity, thermal conductivity, dielectric constant, magnetic
permeability and diffusion coefficient are all mathematically equivalent, thus
our results can also be applied to those cases.  As well as having practical
application, the objective function we consider is convenient from a theoretical
viewpoint due to the cross--property bounds derived by \citet{gibianskyET96}.  

The requirement of isotropy with respect to elasticity has not been seen in
previous multifunctional composite design work, but can be an important criteria
when the directions along which loads will be applied is unknown.  In this case
it is the strength of the weakest direction which is critical.  In particular,
we highlight the anisotropic stiffness properties and weak directions of the
Schwartz P and D structures to justify the isotropy constraint.  As already
noted, these structures were shown to be numerically optimal for maximum bulk
modulus and conductivity by \citet{torquatoET04}.  They possess only cubic
symmetry so are not elastically isotropic and cannot be optimal in the present
context.  

We employ the level-set implementation of topology optimization
\citep{allaireET04,wangET03} to find optimized microstructures.  Microstructure
design problems have not been attempted using this approach; we explore the
capabilities of the method and develop a new way for imposing constraints.

The remainder of the paper is as follows.  Section \ref{problem} outlines the
topology optimization problem, describes the relevant cross--property bounds and
motivates the isotropy constraint.  Section  \ref{algorithm} describes our
computational approach, paying particular attention to how the isotropy
constraint is imposed.  Section \ref{results} presents our results.  Section
\ref{discussion} highlights technical issues and summarizes the results with a
``phase--diagram'' which shows how competition between the multiple objectives
effects the topology of the optimized structures.  Concluding remarks are given
in Section \ref{conclusion}.

\section{Problem outline and motivation}
\label{problem}

\subsection{Problem description}

The phase properties used throughout this paper are $E_1 = 1$, $\nu_1 = 0.3$,
$\sigma_1=0$, $E_2=0$, $\nu_2=0$ and $\sigma_2=1$, where $E_i$ is the Young's
modulus of phase $i$, $\nu_i$ is the Poisson's ratio of phase $i$ and $\sigma_i$
is the conductivity of phase $i$.  Throughout this article the subscript $1$
will refer to the stiff, insulating phase and the subscript $2$ will refer to
the conductive, compliant phase.  These are also referred to as the ``stiff''
and ``conducting'' phases respectively.  

The microstructure is represented by a unit cell cube which may be periodically
extended along each coordinate direction.  All computational results presented
here used a unit cell represented by $40 \times 40 \times 40$ voxels.  We
minimize the objective function 
\begin{equation}
J=-(\frac{\omega_\kappa}{\kappa_1} \kappa^*+ \frac{\omega_\sigma}{\sigma_2}
\sigma^*),
\end{equation}
where $\kappa^*$ and $\sigma^*$ correspond to the effective bulk modulus and
conductivity of the material microstructure and $\omega_\kappa$ and
$\omega_\sigma$ are weights which can be chosen to dictate the relative
importance of the two objectives in our multiobjective design problem.
Throughout this article $\ok$ and $\os$
denote $\frac{\omega_\kappa}{\kappa_1}$ and $\frac{\omega_\sigma}{\sigma_2}$
respectively.  

The microstructure is required to be isotropic with respect to both stiffness
and conductivity.  Specifically, the effective elasticity tensor $A_{ijkl}^*$
must be of the isotropic form
\begin{equation}
\label{eq:isoA}
A_{ijkl}^{*,iso}=\lambda^* \delta_{ij}\delta_{kl}+\mu^*\left( \delta_{ik}
\delta_{jl} + \delta_{il}\delta_{jk}\right),
\end{equation}
where $\mu^*$ is the effective shear modulus of the composite and
$\lambda^*=\kappa^*-\frac{2}{3}\mu^*$ is the other effective Lam\'e constant of
the composite.  
Similarly, the effective conductivity tensor $K_{ij}^*$ of the composite must be
of the isotropic form
\begin{equation}
K_{ij}^{*,iso}=\sigma^* \delta_{ij}.
\label{eq:isoK}
\end{equation}

The volume fraction of each phase is fixed.  Throughout this paper $V_1$ refers
to the volume fraction of the stiff phase and the volume fraction of the
conductive phase is $V_2=1-V_1$.    

Details regarding finding effective properties and imposing the required
constraints are left to Section~\ref{algorithm}.

\subsection{Cross--property bounds}
\label{cross-property}

Cross--property bounds for conductivity and bulk modulus rigorously restrict the
effective properties of a composite to be in some allowable region of the
$\sigma^*-\kappa^*$ plane. They are used here to demonstrate the
near--optimality of our optimized microstructrures.  

The tightest known cross-property bounds between bulk modulus and conductivity
for three dimensional two--phase isotropic or cubic--symmetric composites were
derived by \citet{gibianskyET96}.  
For the particular phase properties used in this paper 
the set of all possible $(\sigma^*,\kappa^*)$ pairs is bounded by the two lines
between the points
\begin{equation}
\{(0,0),(\sigma_{HS},0)\} \quad \mathrm{and} \quad \{(0,0),(0,\kappa_{HS})\} 
\end{equation}
and by the hyperbola 
parameterized by:
\begin{eqnarray}
\sigma^* & = & (1-\gamma) \sigma_{HS}-\frac{\gamma (1-\gamma)
\sigma_{HS}^2}{\gamma\sigma_{HS}-(1-V_1)\sigma_2},\nonumber \\
\kappa^* & = & \gamma\kappa_{HS}-\frac{\gamma(1-\gamma)
\kappa_{HS}^2}{(1-\gamma) \kappa_{HS} -V_1 \kappa_1} ,
\label{eq:hyp}
\end{eqnarray}
where $\gamma \in [0,1]$. 
$\sigma_{HS}$ and $\kappa_{HS}$ denote the upper Hashin--Shtrikman bounds
on the effective conductivity \citep{hashinET62} and the effective bulk modulus
\citep{hashinET63}.

\subsection{Schwartz primitive and diamond structures}

Composites with Schwartz P and D minimal surfaces as the phase interface must
have effective properties which satisfy the cross--property bounds above for
$V_1=\frac{1}{2}$.  \citet{torquatoET04} showed via numerical calculations that
Schwartz P and D structures have properties which lie on 
a particular point on the cross--property bounds for any ill--ordered phase
properties.
For the particular phase properties considered in this paper, the Schwartz P and
D structures have effective properties numerically equal to
$(\sigma^*_U,\kappa^*_U) = (\frac{1}{3},\frac{10}{63})$.  

\begin{figure}[!htb]
\begin{center}
\includegraphics[width=0.95\textwidth]{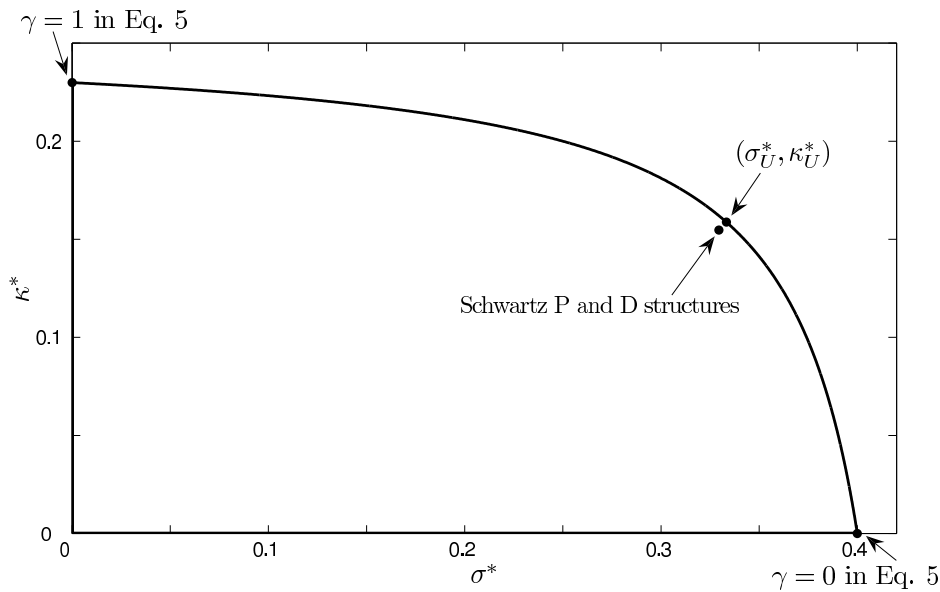}\\
\end{center}
  \caption{Effective properties of the $64,000$ voxel Schwartz P and D
structures, the theoretical optimal point $(\sigma^*_U,\kappa^*_U)$ and the
relevant cross--property bounds.  The effective properties must
lie within the region bounded by hyperbola shown and the $\sigma^*=0$
and $\kappa^*=0$ axes.}
  \label{fig:bounds_pd}
\end{figure}

Fig.~\ref{fig:bounds_pd} shows the cross--property bounds, the optimal point
$(\sigma^*_U,\kappa^*_U)$, and the calculated properties of $40 \times 40 \times
40$ voxel approximations to the Schwartz P and D structures for our particular
phase properties.  As expected, our calculated points do not exactly coincide
with the theoretically optimal point due to the use of only $64,000$ voxels.  

\begin{figure}[!htb]
\begin{center}
\begin{minipage}{5.5cm}
\includegraphics[width=5cm]{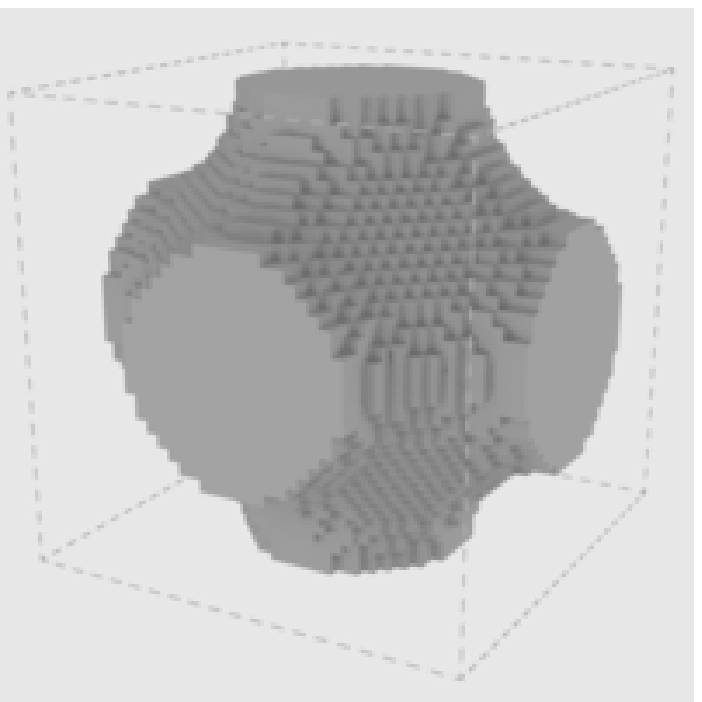}
\end{minipage}
\begin{minipage}{5.5cm}
\vspace{0.5cm}
\includegraphics[width=6cm]{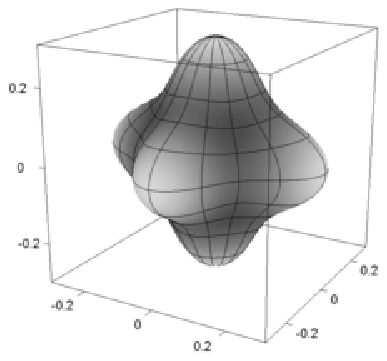}
\vspace{-0.5cm}
\end{minipage}
\end{center}
  \caption{The conductive phase of the $64,000$ voxel approximation to a
Schwartz P structure (left), and the directional dependence of its effective
Young's modulus (right). }
  \label{fig:p_surf}
\end{figure}

\begin{figure}[!htb]
\begin{center}
\begin{minipage}{5.5cm}
\includegraphics[width=5cm]{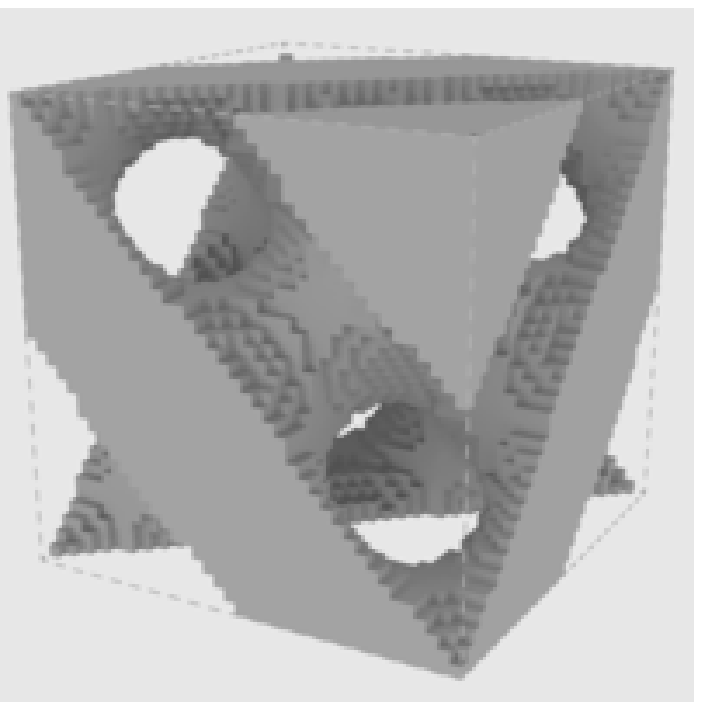}
\end{minipage}
\begin{minipage}{5.5cm}
\includegraphics[width=6cm]{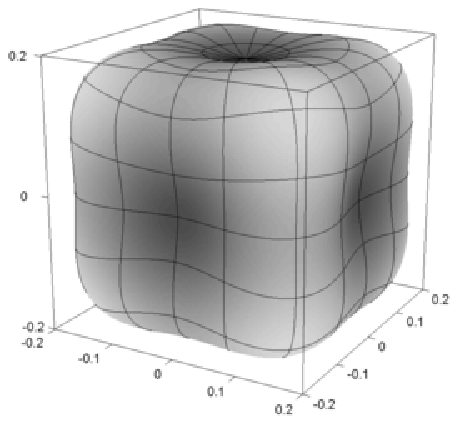}
\vspace{-0.5cm}
\end{minipage}
\end{center}
  \caption{The conductive phase of the $64,000$ voxel approximation to a
Schwartz D structure (left), and its calculated effective Young's modulus
for all directions (right). }
  \label{fig:d_surf}
\end{figure}

The conductive phase of the approximate Schwartz P and D structures is shown in
Figs.~\ref{fig:p_surf} and \ref{fig:d_surf} respectively.   These figures also
show the directional dependence of the effective Young's modulus for the
Schwartz P and D structures.  

The effective Young's modulus $E^*$ in a particular direction can be measured by
loading the material uniaxially along that direction.  $E^*$ is then the value
of the applied stress divided by the resulting strain as measured along the
loaded direction.  
Once the effective elasticity tensor for the composite has been calculated its
Young's modulus can be readily determined for any direction.  We define
$E_{min}^*$ and $E_{max}^*$ as the value of the effective Young's modulus along
the directions for which it is smallest and largest respectively.  

Table~\ref{table:minsurfs} tabulates the calculated properties of $64,000$
voxel approximations of the Schwartz P and D structures.  

\begin{table}[!htb]
  \caption{Effective properties of the approximate Schwartz P and D structures
pictured in Figs.~\ref{fig:p_surf} and \ref{fig:d_surf}.  }
\label{table:minsurfs}
\begin{tabular*}{\textwidth}{@{\extracolsep{\fill}}c c c c c c}
\hline
Structure & $\sigma^*$ & $\kappa^*$ & $\aniso$ & $E^*_{min}$ & $E^*_{max}$ \\
\hline
Schwartz P & 0.3288 & 0.1542 & 0.1925 & 0.1918 & 0.3040 \\
\hline
Schwartz D & 0.3302 & 0.1551 & 0.1635 & 0.1800 & 0.2795 \\
\hline
\end{tabular*}
\begin{minipage}{\textwidth}
\vspace{0.1cm}
\emph{Notes:} $\aniso$ is the anisotropy of the structure, as defined in Section
\ref{algorithm}.  \end{minipage}
\end{table}

\subsection{Motivation for the isotropy constraint}

Figs.~\ref{fig:p_surf} and \ref{fig:d_surf} clearly show the anisotropic
stiffness properties of the Schwartz P and D structures.  This is also
demonstrated by the large difference between $E_{min}^*$ and $E_{max}^*$ for the
two structures, as tabulated in Table~\ref{table:minsurfs}.  In many engineering
applications the presence of weak directions would not be favourable.  We note
that the Schwartz P and D structures have cubic symmetry and therefore are
isotropic with respect to conductivity.  Structures without cubic symmetry 
have directions with low conductivity, this would also not be desirable
in some cases.  An isotropy constraint is an obvious suggestion to ensure that
the Young's modulus and conductivity are the same in all directions.  In such a
case the picture corresponding to the right hand picture in
Figs.~\ref{fig:p_surf} and \ref{fig:d_surf} would be a sphere with radius $E^*$.

More precisely, the two following intuitive result holds.  

\subsubsection{Result}

Given isotropic and anisotropic microstructures with the same average value of
any directional property $X^*$, the isotropic microstructure has the largest
$X_{min}^*$.  

\subsubsection{Proof}

The property $X^*$ is averaged over all directions for the anisotropic
microstructure to give $\bar{X^*}$:
\begin{equation}
\bar{X^*}=\frac{1}{4 \pi}\int_{\theta=0}^{2\pi}\int_{\phi=0}^{\pi}
X^*(\theta,\phi) \sin(\phi) \dd \phi \dd \theta,
\end{equation}
where $\theta$ and $\phi$ are the standard azimuthal and polar angles
respectively.  Since $\bar{X^*}$ is the average of $X^*$ for all directions and
the microstructure is not isotropic, there must exist a direction for which
$X^*<\bar{X^*}$.  Thus $X_{min}^{*,aniso}<\bar{X^*}$.  

Since the isotropic microstructure has this same $\bar{X^*}$ and is isotropic,
$X^{*,iso}=\bar{X^*}$ for all directions.  Obviously $X_{min}^{*,iso}=\bar{X^*}$.  

We obtain $X_{min}^{*,aniso}<\bar{X^*}=X_{min}^{*,iso}$, proving the result.

\section{Topology optimization algorithm}
\label{algorithm}

\subsection{Calculation of effective properties}
\label{properties}

We use the standard elastic homogenization approach, for example see
\citet{garbocziET95} and the references therein, and choose to write it simply
as
\begin{equation} \label{eq:elas}
\sigma^*_{ij}=A_{ijkl}^*\epsilon^*_{kl}, 
\end{equation}
where $\sigma_{ij}^*$ represents the effective stress tensor (not to be confused
with $\sigma^*$ which represents the effective conductivity) and
$\epsilon^*_{kl}$ represents the global strain tensor.  The usual summation
convention is used.  The effective elasticity tensor $A_{ijkl}^*$ can be
calculated by applying six independent strain fields corresponding to the six
independent components of the symmetric strain tensor.  These strains are
applied to the unit cell with periodic boundary conditions.  The resulting
stresses are found using the finite element method and are averaged over the
voxels to give the components of the elasticity tensor.

Similarly for the conductivity case, 
\begin{equation}
J^*_i=K_{ij}^*E^*_j,
\end{equation}
where $J^*_i$ is the effective current vector and $E^*_i$ is the global applied
electric field.  The effective conductivity tensor $K_{ij}^*$ is calculated by
applying three independent electric fields to the unit cell with periodic
boundary conditions.  The electric currents are computed by solving Laplace's
equation with the finite element method and are averaged over the voxels in the
finite element mesh to give the components of the conductivity tensor.  

We use the following definitions of the effective bulk modulus and shear
modulus:
\begin{eqnarray}
\kappa^* &= &\frac{1}{9}A^*_{iijj}, \label{eq:effkappa} \\ 
\mu^* & = & \frac{1}{20}\left(A^*_{ijij}+A^*_{ijji}\right) -
\frac{1}{30}A^*_{iijj} \label{eq:effmu},
\end{eqnarray}
which are invariants of the effective elasticity tensor and are certainly
correct for the isotropic case in Eq.~\ref{eq:isoA}.

Similarly, the following is our definition of the effective conductivity
\begin{equation}
\sigma^*=\frac{1}{3}K_{ii}^*. \label{eq:effsigma}
\end{equation}

\subsection{Measuring anisotropy}

We measure elastic anisotropy as the ``distance'' between the calculated
$A_{ijkl}^*$ and the ``nearest'' isotropic elasticity tensor $A_{ijkl}^{*,iso}$:
\begin{equation}
\aniso_{mech}=\sqrt{\frac{\left(A_{ijkl}^*-A_{ijkl}^{*,iso}\right)
\left(A_{ijkl}^*-A_{ijkl}^{*,iso}\right)}{A_{\bar{i}\bar{j}\bar{k}\bar{l}}^{*,iso}
A_{\bar{i}\bar{j}\bar{k}\bar{l}}^{*,iso}}},
\end{equation}
with $A_{ijkl}^{*,iso}$ given in Eq. \ref{eq:isoA} using effective properties
from Eqs. \ref{eq:effkappa} and \ref{eq:effmu}.  The denominator can be
simplified to $9 \kappa^{*2}+20 \mu^{*2}$.  

Similarly, we measure conductive anisotropy as
\begin{equation}
\aniso_{cond}=\sqrt{\frac{\left(K_{ij}^*-K_{ij}^{*,iso}\right)
\left(K_{ij}^*-K_{ij}^{*,iso}\right)}{K_{\bar{i}\bar{j}}^{*,iso}
K_{\bar{i}\bar{j}}^{*,iso}}},
\end{equation}
where $K_{ij}^{*,iso}$ is as in Eq. \ref{eq:isoK} with $\sigma^*$ defined in
Eq. \ref{eq:effsigma}.  The denominator simplifies to $3 \sigma^{*2}$.

As an overall anisotropy measure these two individual anisotropies are summed to
give 
\begin{equation}
\aniso=\aniso_{mech}+\aniso_{cond}.
\end{equation}

\subsection{Isotropy constraints}

The numerators of $\aniso_{mech}^2$ and $\aniso_{cond}^2$ are sums of squares.
To impose $\aniso=0$ a constraint is introduced for each term in the sum.  For
example, in the conductivity case the six constraints are $C_p, p\in
\{1,\dots,6\}$, where
\begin{eqnarray}
\sqrt{3} \sigma^* C_1 & = & K_{11}^*-\sigma^*,\nonumber \\
\sqrt{3} \sigma^* C_2 & = & K_{22}^*-\sigma^*,\nonumber \\
\sqrt{3} \sigma^* C_3 & = & K_{33}^*-\sigma^*,\nonumber \\
\sqrt{3} \sigma^* C_4 & = & \sqrt{2}K_{12}^*,\nonumber \\
\sqrt{3} \sigma^* C_5 & = & \sqrt{2}K_{13}^*,\nonumber \\
\sqrt{3} \sigma^* C_6 & = & \sqrt{2}K_{23}^*.
\end{eqnarray} 
Only five of these are independent but we work with all six to explicitly retain
symmetry in the numerical calculations.  The factors of $\sqrt{2}$ are included
in $C_4$, $C_5$ and $C_6$ on the right hand side and the factors of $\sqrt{3}
\sigma^*$ are included on the left hand side to give
$\aniso_{cond}^2=\sum_{p=1}^{6}C_p^2$.  

Similarly $21$ constraints $C_{p},p\in\{7,\dots,27\}$ can be written down for the
elasticity case, $19$ of which are independent.  They are also normalized so
that summing and squaring them gives $\aniso_{mech}^2$.

\subsection{Level--set method for topology optimization}

Topological optimization is implemented using the level--set approach formalized
by \citet{allaireET04} and \citet{wangET03} and based on earlier work by
\citet{osherET01} and \citet{sethianET00}.  The reader is referred to
\citet{allaireET04} for standard details of the level--set approach; only
significant deviations from the standard method are mentioned here.  Note that
the topological derivative, as used for example by \citet{allaireET05a}, is not
used in the work presented here.  We have found that, as noted by
\citet{allaireET04}, topological changes are readily achieved using the
level--set method in three dimensions without the topological derivative,
rendering it unnecessary.  


In our implementation the level--set function is initialized using an
approximation to the signed distance function, found by finding the nearest
voxel of opposite phase with respect to the standard metric on $\mathbb{R}^3$.
This initialization procedure is also used to reinitialize the level--set
function at each iteration of the algorithm.


The Hamilton--Jacobi evolution equation is solved via a standard upwind scheme
and the time--step for the numerical evolution must be less than that given by
the Courant--Friedrichs--Lewy (CFL) condition to ensure numerical stability
\citep{sethian99}.  We choose to use a time--step 1\% of the CFL value and do
many level--set evolutions per iteration of the optimization algorithm,
typically between 25 and 100.  The number of level--set evolutions per iteration
is gradually reduced as the optimization converges to a local minimum.

The shape derivatives of the effective elasticity and conductivity tensor
components $A_{ijkl}^*$ and $K_{ij}^*$ are readily computed from well--known
shape derivatives \citep[see][and the references therein]{allaireET04} by
utilizing the superposition principle for strains and stresses in linear
elasticity and electric fields and currents in conductivity.  From these it is
straightforward to calculate the shape derivatives of the
constraints\footnote{The scaling factors of the constraints from the
denominators of the anisotropy measures are not considered as variables but as
constant scaling factors for the purposes of this calculation.} and of the
effective properties $\kappa^*$ and $\sigma^*$ to give the shape derivative of
the objective function.  These are of the form
\begin{eqnarray}
\frac{\dd J}{\dd \Omega}(\theta) &= &-\int_{\partial \Omega}
\theta_i n_i v, \nonumber \\
\frac{\dd C_p}{\dd \Omega}(\theta) &= &-\int_{\partial \Omega}
\theta_i n_i v_p \quad p\in\{1,\dots,27\}.
\end{eqnarray}
Here $\Omega$ is the region occupied by the stiff phase.  It is being deformed
by the map $x_i\rightarrow x_i+\theta_i$, so the left hand side of each equation
above is the standard shape derivative.   The boundary of $\Omega$ is $\partial
\Omega$, with outward normal $n_i$.  The quantities $v$ and $v_{p}$ are
called the shape sensitivities of $J$ and $C_{p}$.  

\subsection{Imposing the isotropy constraints}

In the situation with no constraints, $\theta_i$ is often chosen to be
$\theta_i=vn_i$, and the normal velocity $\mathcal{V}$ of the phase interface is
chosen as $\mathcal{V}=\thdn=v$.  This implements a steepest descent type algorithm
under evolution with the Hamilton--Jacobi equation.    A modification of this
for the constrained case is outlined below.  A more detailed presentation is
given in \citet{wilkinsET07}.  

At any generic iteration of the algorithm, the constraints will be
nonzero.  To reduce both the objective function and the constraints we
can require:
\begin{eqnarray}
\int_{\pl\Omega}\thdn v & = & \mbox{maximum possible},
\label{dsec.sdf}
\\
\int_{\pl\Omega}\thdn v_{p} & = & r C_{p} \ \quad p\in\{1,\dots,27\}, 
\label{const.res} 
\end{eqnarray}
where $r$ dictates the rate of exponential decay of the constraints.

The shape sensitivities are functions defined on $\pl\Omega$ and may be thought of as
elements of a Hilbert space.  In the following, the notation $||\cdot||$ and
$\lan \cdot,\cdot\ran$ is short hand for the norm and inner product on $\pl\Omega$:
$||v_p||^2=\int_{\pl\Omega}v_p^2 $, $\lan v_{p},v_{q}\ran =
\int_{\pl\Omega}v_{p}v_{q}$.  In the finite element implementation
$\int_{\pl\Omega} v_{p} v_{q}$ is approximated by a sum over all the boundary
voxels of the product $ v_{p} v_{q}$.  A voxel is determined as a boundary voxel
if any of its 26 neighbours (including those across periodic boundaries) are of
the opposite phase.  

Linearly dependent $\{v_p\}$ are removed from the set.  We use the Gram-Schmidt
procedure to build a mutually orthogonal set $\{\bar{v}_{p}:
p\in\{1,\dots,24\}\}$ from $\{v_{p}\}$ which spans the constraint shape
sensitivities.
From this we can form the projection operator $P$ which projects vectors in
$\pl\Omega$ onto the space which will leave the constraints invariant.

To implement Eq.~\ref{const.res} we take
\begin{equation}
\mathcal{V}=\thdn = \sqrt{1-\sum_{p=1}^{24}\al_{p}^{2}}\frac{Pv}{||Pv||} +
\sum_{p=1}^{24}\al_{p}\frac{\bar{v}_{p}}{||\bar{v}_{p}||} \ ,
\label{eqn.thdn.good}
\end{equation}
where the $\al_{p}$ are real numbers which are chosen to solve 
the lower-diagonal system
\begin{equation}
\left(
\begin{array}{c}
r C_{1} \\ r C_{2} \\ r C_{3} \\ \vdots \\ rC_{24}
\end{array}
\right)
= \left(
\begin{array}{ccccc}
||\bar{v}_{1}|| & 0 & 0 & \ldots & 0 \\
\frac{\lan\bar{v}_{1}, v_{2}\ran}{||\bar{v}_{1}||} & ||\bar{v}_{2}|| & 0 & \ldots
& 0 \\
\frac{\lan\bar{v}_{1}, v_{3}\ran}{||\bar{v}_{1}||} & 
\frac{\lan\bar{v}_{2}, v_{3}\ran}{||\bar{v}_{2}||} & 
||\bar{v}_{3}|| & \ldots & 0 \\
\vdots & \vdots & \vdots & \ddots & 0 \\
\frac{\lan\bar{v}_{1}, v_{24}\ran}{||\bar{v}_{1}||} &
\frac{\lan\bar{v}_{2}, v_{24}\ran}{||\bar{v}_{2}||} &
\frac{\lan\bar{v}_{3}, v_{24}\ran}{||\bar{v}_{3}||} &
\ldots & ||\bar{v}_{24}||
\end{array}
\right)
\left(
\begin{array}{c}
\al_{1} \\ \al_{2} \\ \al_{3} \\ \vdots \\ \al_{24}
\end{array}
\right).
\label{giving.alphas}
\end{equation}
In essence, this has decomposed $\thdn$ into the sum of two parts. The first
part is orthogonal to the shape sensitivities.  The second part is a linear
combination of the shape sensitivities.  The former has been chosen to decrease
the objective function as in Eq.~(\ref{dsec.sdf}), while the latter has been
chosen to decrease the constraints as in Eq.~(\ref{const.res}).  

The parameter $r$ is chosen to ensure \mbox{$1-\sum_{p}\al_{p}^{2}\geq 0$} and
$\sum_{p}\al_{p}^{2}\geq \al_{min}^2$.  Changing $\al_{min}^2$ effects how
strongly the algorithm projects onto the constraints.  Typically we choose
$\al_{min}^2=0.1$.  


\subsection{Imposing the volume constraint}

The volume constraint is imposed via the Karush--Khun--Tucker technique
\citep{karush39,kuhnET51}.  In our setting this technique involves calculating the
expected volume change under evolution and if necessary correcting the shape
derivative to ensure the volume change will not make the phase volume fractions
deviate further from their required values.  This type of technique has
been used with the level--set method for topology optimization previously
\citep{wangET06}.

\subsection{Velocity extension via smoothing}

The above method for the isotropy constraints calculates the normal
velocity $\mathcal{V}$ at all boundary voxels.  The normal velocity is set to zero for all
non--boundary voxels.  

One issue with the level--set method of topology optimization is that the
velocity must be extended away from the boundaries to give evolution of the
structure.  Frequently this is addressed by using an ersatz material approach,
whereby material properties of the weak phase are set to some small nonzero
value instead of to zero, see for example \citet{allaireET04}.  We choose to apply a
smoothing to the velocities via the convolution
\begin{equation}
\mathcal{V}_{i,j,k} \gets \frac{1}{8}(2 \mathcal{V}_{i,j,k}
+\mathcal{V}_{i-1,j,k} +\mathcal{V}_{i+1,j,k} +\mathcal{V}_{i,j-1,k}
+\mathcal{V}_{i,j+1,k} +\mathcal{V}_{i,j,k-1} +\mathcal{V}_{i,j,k+1}),
\end{equation}
where the subscripts refer to indices of the voxels.  This convolution is
applied repeatedly until Hamilton--Jacobi evolution will result in a geometric
change of the structure.  We note that this smoothing operation allows us to use
only two phases during the optimization: no intermediate densities are
required.

\section{Results}
\label{results}

\begin{figure}[!htb]
\begin{center}
\includegraphics[width=3.1cm]{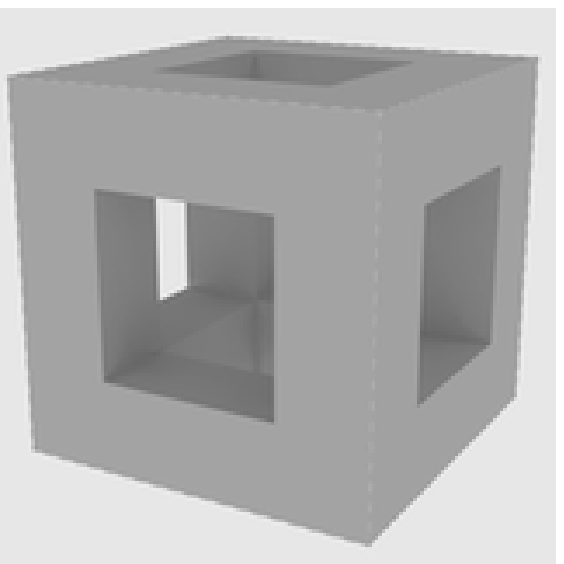}~
\includegraphics[width=3.1cm]{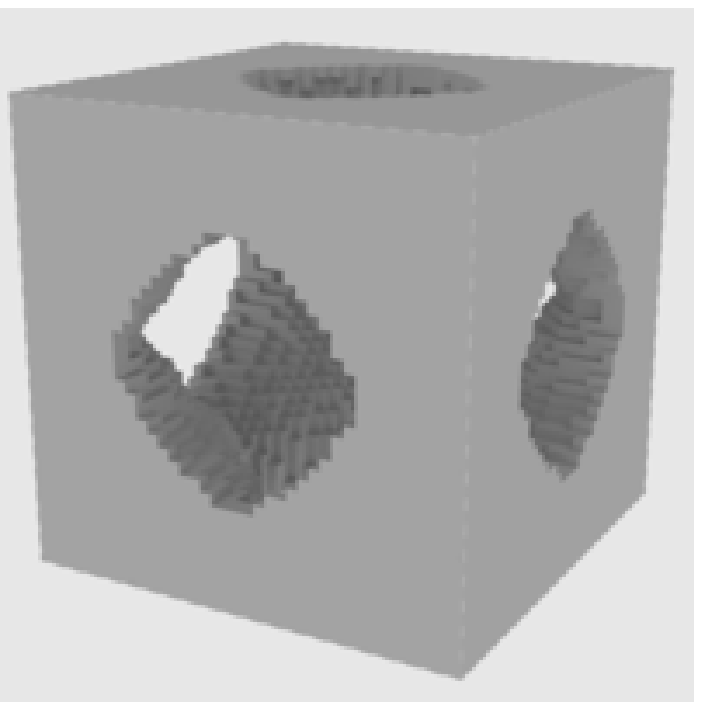}~
\includegraphics[width=3.1cm]{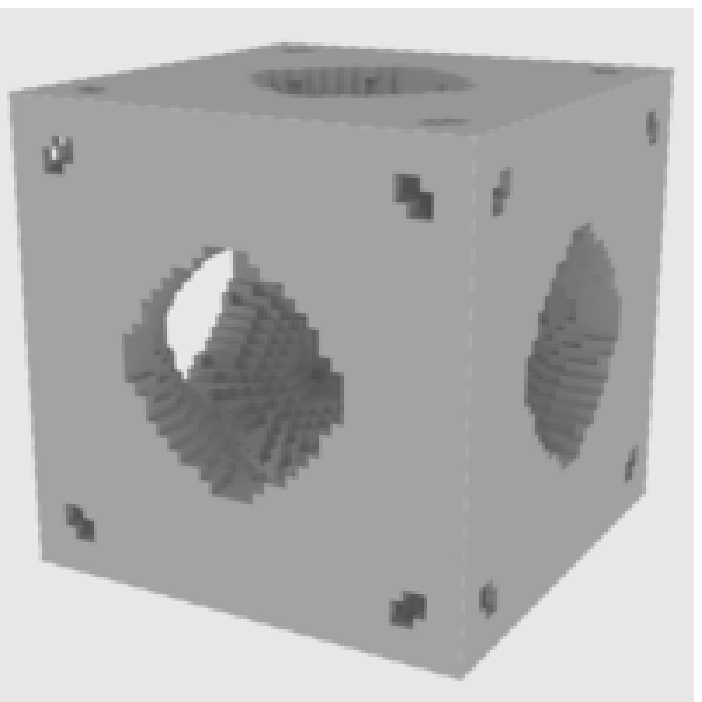}~
\includegraphics[width=3.1cm]{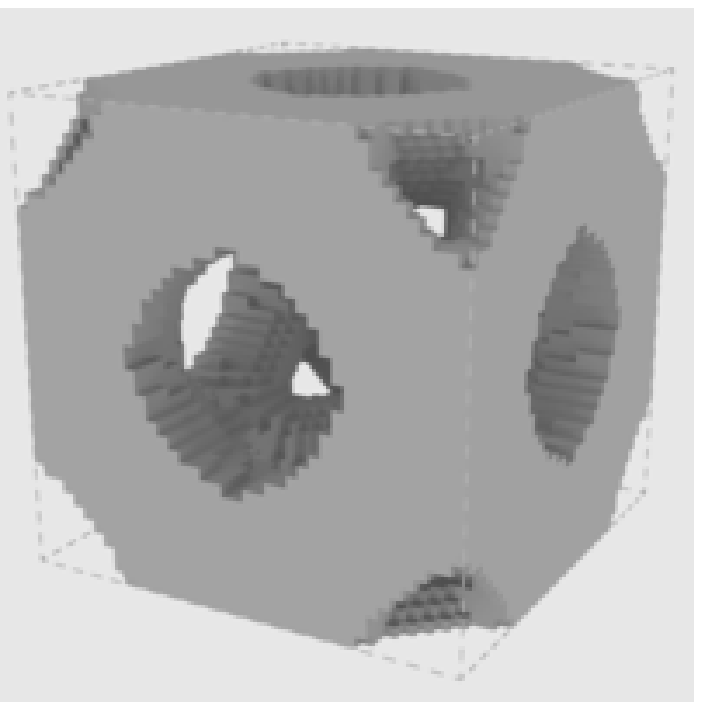}\\
\includegraphics[width=3.1cm]{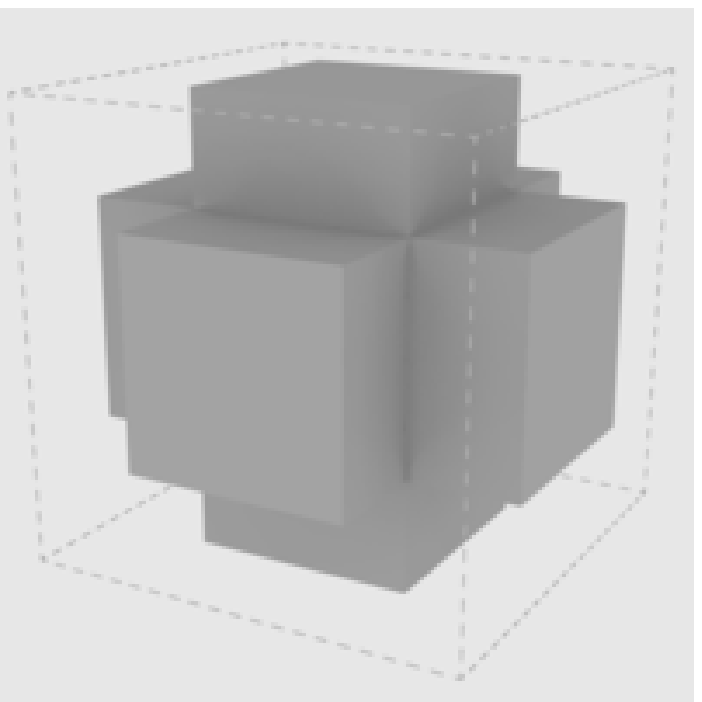}~
\includegraphics[width=3.1cm]{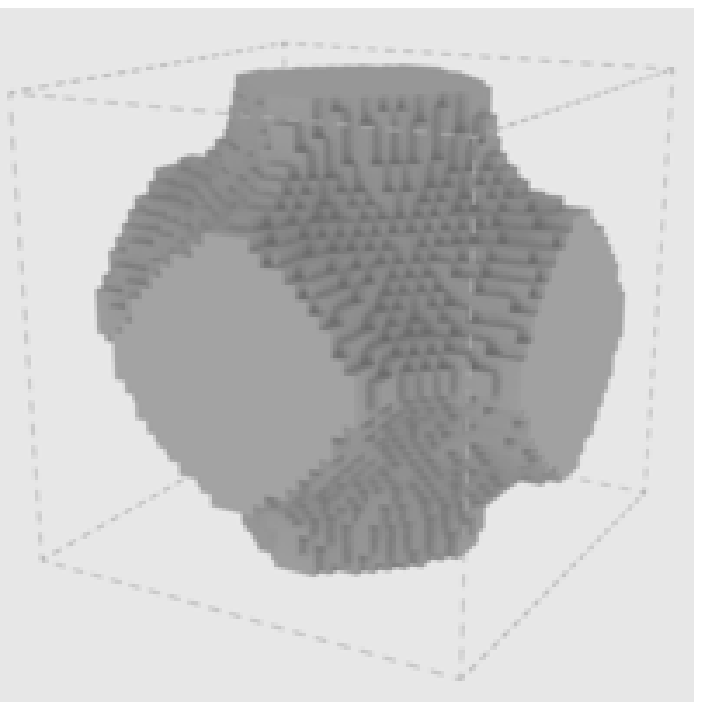}~
\includegraphics[width=3.1cm]{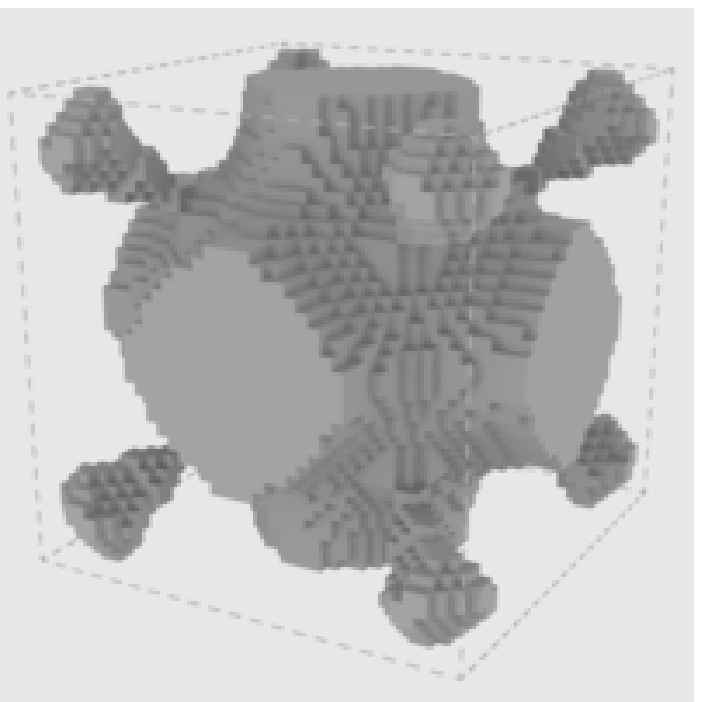}~
\includegraphics[width=3.1cm]{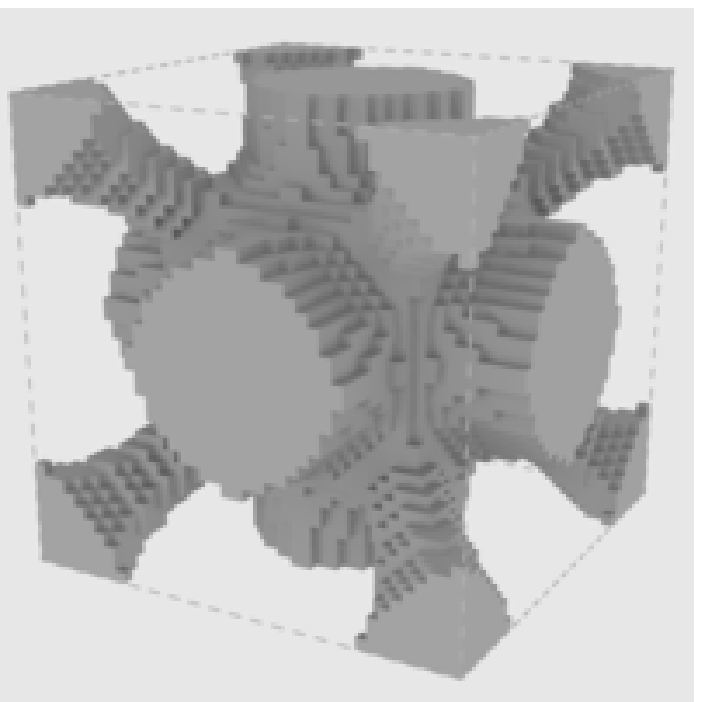}\\
(a) \hspace{2.5cm} (b) \hspace{2.5cm} (c) \hspace{2.5cm} (d)
\end{center}
  \caption{An example optimization history with $(\ok,\os)=(1,\frac{1}{2})$.  The four
columns are (a) the initial unit cell, (b) iteration 20, (c) iteration 55 and
(d) the optimized unit cell.  In each case the top and bottom picture show the stiff phase
and conductive phase respectively.}
  \label{fig:egoptim}
\end{figure}

Any initial unit cell is suitable to start the optimization process provided
the initial periodic material has nonzero stiffness and nonzero conductivity in
all directions.  All of the optimized structures presented are optimized from the initial
unit cell shown in Fig.~\ref{fig:egoptim}(a).  

The chosen initial unit cell has two nice properties.  First, the two phases
have the same topology and geometry, meaning that there is no initial bias
towards a particular property.  Although the two phases of the initial unit cell
look different in Fig.~\ref{fig:egoptim}(a), a shift of either phase along each
coordinate direction by half the unit cell edge length demonstrates that the two
phases of the initial microstructure are geometrically identical.  Second, the
cell has simple cubic symmetry.  Our computational approach retains this
throughout the optimization and $\aniso_{cond}=0$ at each iteration.  Isotropy
with respect to elasticity is more difficult to achieve.  

The choice of internal parameters in the algorithm effects the outcome of the
optimization process.  The differences are often insignificant and experience
with the algorithm allows the user to find a small set of internal parameters to
use on each optimization problem.  Results presented reflect the optimized
structure with the best objective function value from optimizations with various
choices of internal parameters.  It is not surprising that other inferior
structures can be reached with different parameter choices: the
level--set method converges to a local minimum and we do not know that a
single global minimum exists.

\subsection{Equal phase volume fractions}

First we consider the design of unit cells with an equal volume fraction for each
phase, i.e., $V_1=\frac{1}{2}=V_2$.  Fig.~\ref{fig:egoptim} shows an initial
structure, two intermediate structures and the optimized structure for
$(\ok,\os)=(1,\frac{1}{2})$.  Fig.~\ref{fig:egoptim} highlights that no
intermediate densities are used at any time in the optimization.  

The material properties for the four microstructures shown in
Fig.~\ref{fig:egoptim} are given in Table~\ref{table:egoptim}.  Note that the
volume constraint and the isotropy constraint are not satisfied during the
optimization process.  The errors on these constraints are small for the
optimized structure (and for other optimized structures presented).  We have
seen that small geometric changes  can produce large percentage changes in
$\aniso$ when it is small, with corresponding very small changes in $V_1$, $V_2$
and the objective function.  Thus we consider a value of $\aniso<0.005$ to be
small enough for the isotropy constraint to be satisfied.  Also, a value of
$|V_1^{required}-V_1^{actual}|<0.0005$ is considered good enough given that the
unit cells are represented with only $64,000$  voxels without intermediate
densities.  

Particularly of interest in Table~\ref{table:egoptim} is how the minimum and
maximum Young's moduli $E_{min}^*$ and $E_{max}^*$ change throughout the
optimization.  As the optimization progresses and the anisotropy $\aniso$
decreases, the difference between the maximum and minimum Young's moduli also
decreases.  For the optimized structure we see that the maximum and minimum
Young's moduli are less than 1\% apart, highlighting the effectiveness of the
isotropy constraint.  This is the case for all optimized structures presented
and therefore $E^*=\frac{9\kappa^*\mu^*}{3 \kappa^*+\mu^*}$ is tabulated in the
remainder of the paper.

\begin{table}[!htb]
  \caption{Effective properties of the four unit cells from the optimization
history shown in Fig.~\ref{fig:egoptim}.  }
  \label{table:egoptim}
\begin{tabular*}{\textwidth}{@{\extracolsep{\fill}}c c c c c c c c}
\hline
Case & $J$ & $\sigma^*$ & $\kappa^*$ & $V_1$ & $\aniso$ & $E^*_{min}$ & $E^*_{max}$ \\
\hline
(a) & -0.3016 & 0.3170 & 0.1432 & 0.5000 & 0.2826 & 0.1444 & 0.2951 \\
\hline
(b) & -0.3063 & 0.3261 & 0.1432 & 0.5109 & 0.1806 & 0.1901 & 0.2885 \\
\hline
(c) & -0.3170 & 0.2987 & 0.1676 & 0.5194 & 0.0909  & 0.2322  & 0.2902 \\
\hline
(d) & -0.3187 & 0.3226 & 0.1574 & 0.5003 & 0.0029 &  0.2373 & 0.2389 \\
\hline
\end{tabular*}
\end{table}

\begin{figure}[!htb]
\begin{center}
\includegraphics[width=3.1cm]{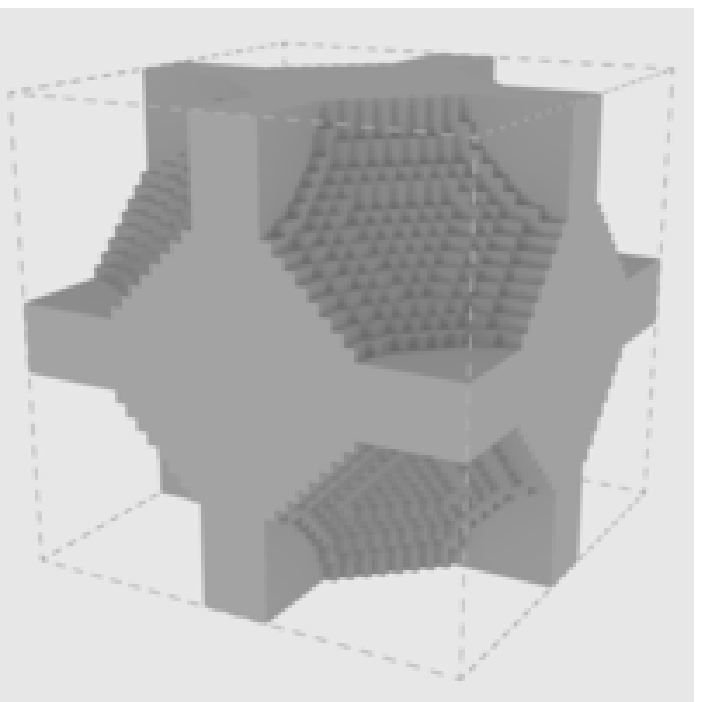}~
\includegraphics[width=3.1cm]{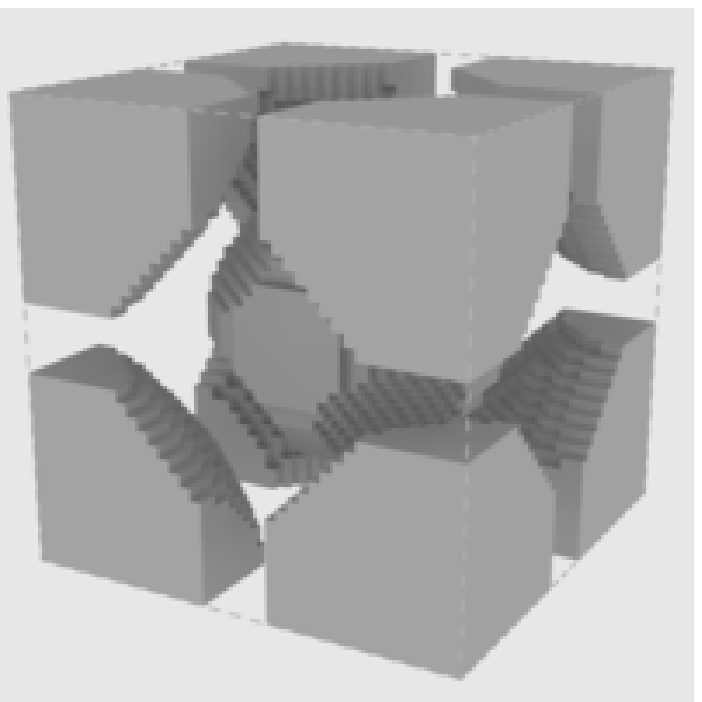}~\hspace{0.5cm}~
\includegraphics[width=3.1cm]{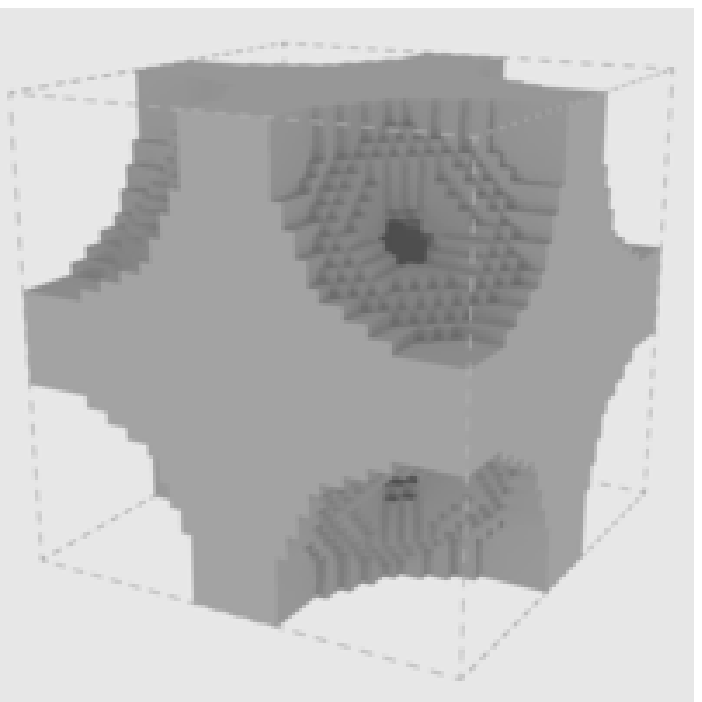}~
\includegraphics[width=3.1cm]{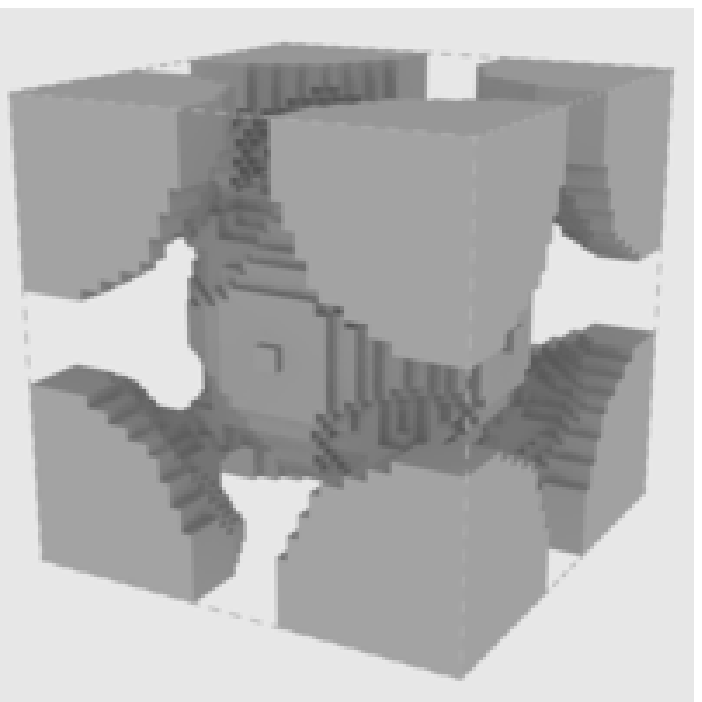}\\
(a) \hspace{6.5cm} (b) \\
\includegraphics[width=3.1cm]{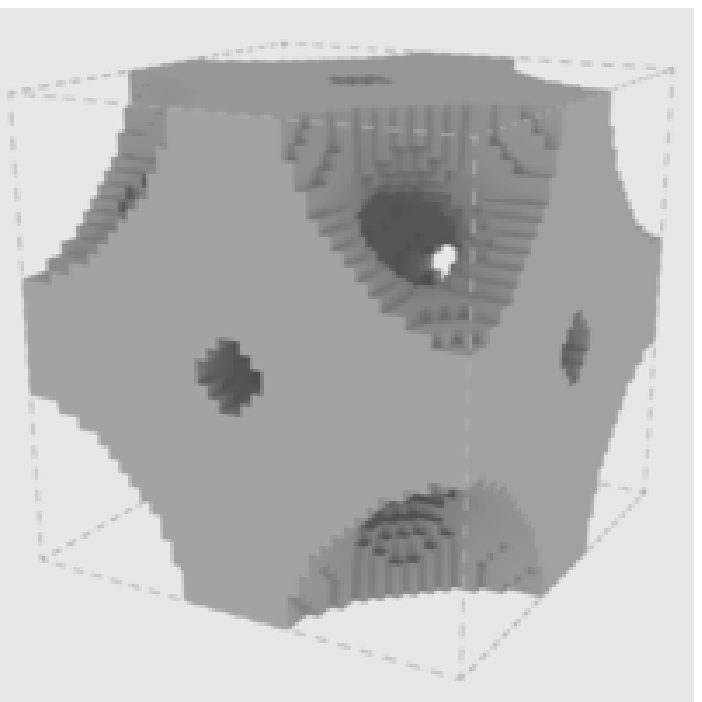}~
\includegraphics[width=3.1cm]{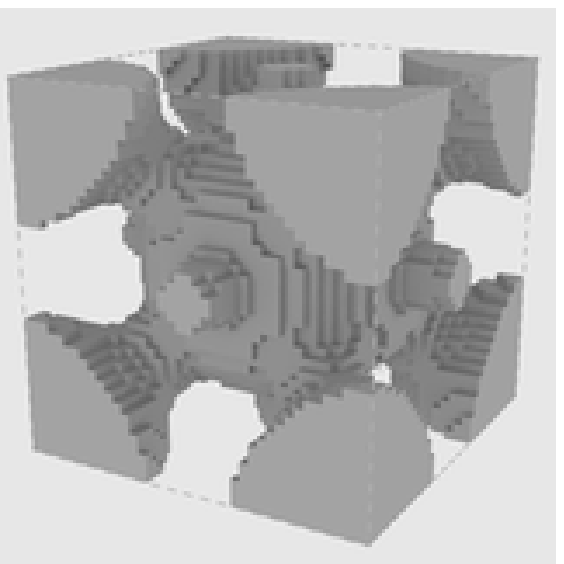}~\hspace{0.5cm}~
\includegraphics[width=3.1cm]{final_struc_solid_2_1.eps}~
\includegraphics[width=3.1cm]{final_struc_void_2_1.eps}\\
(c) \hspace{6.5cm} (d) \\
\includegraphics[width=3.1cm]{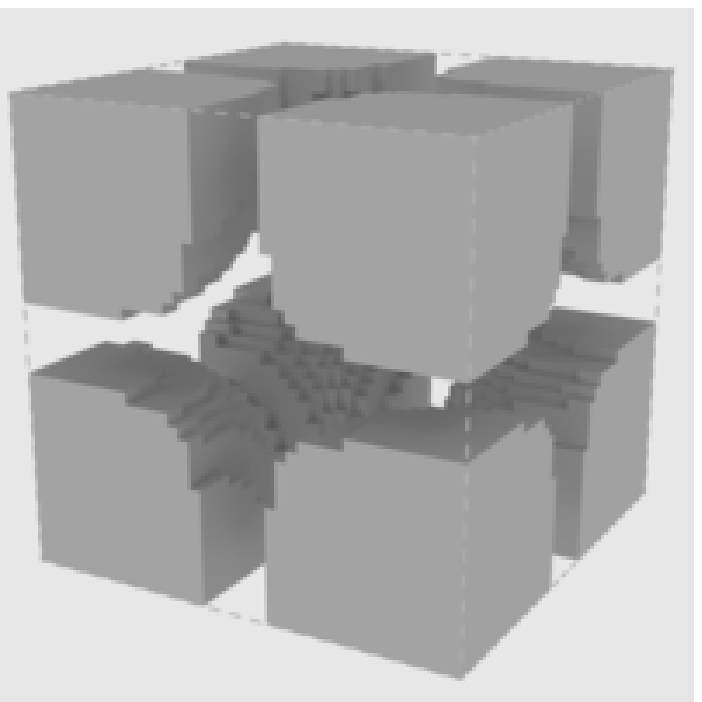}~
\includegraphics[width=3.1cm]{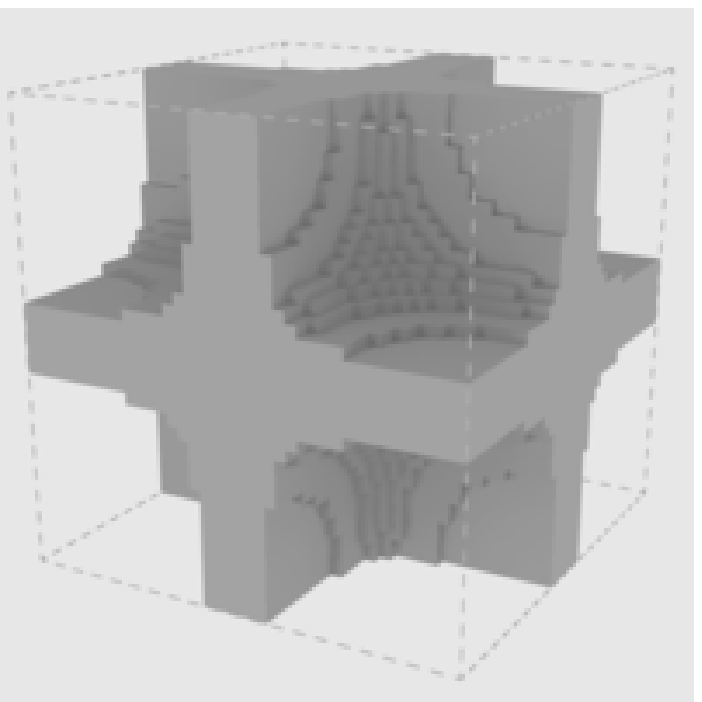}\\
(e)
\end{center}
  \caption{Optimized unit cells with $V_1=\frac{1}{2}$ and different weighting
schemes in the objective function: (a) $(\ok,\os)=(1,0)$, (b)
$(\ok,\os)=(1,\frac{1}{10})$, (c) $(\ok,\os)=(1,\frac{1}{6})$, (d)
$(\ok,\os)=(1,\frac{1}{2})$, (e) $(\ok,\os)=(0,1)$.  In each case the left and
right picture show the stiff phase and conductive phase respectively. }
  \label{fig:50_50}
\end{figure}

\begin{figure}[!htb]
\begin{center}
\includegraphics[width=0.95\textwidth]{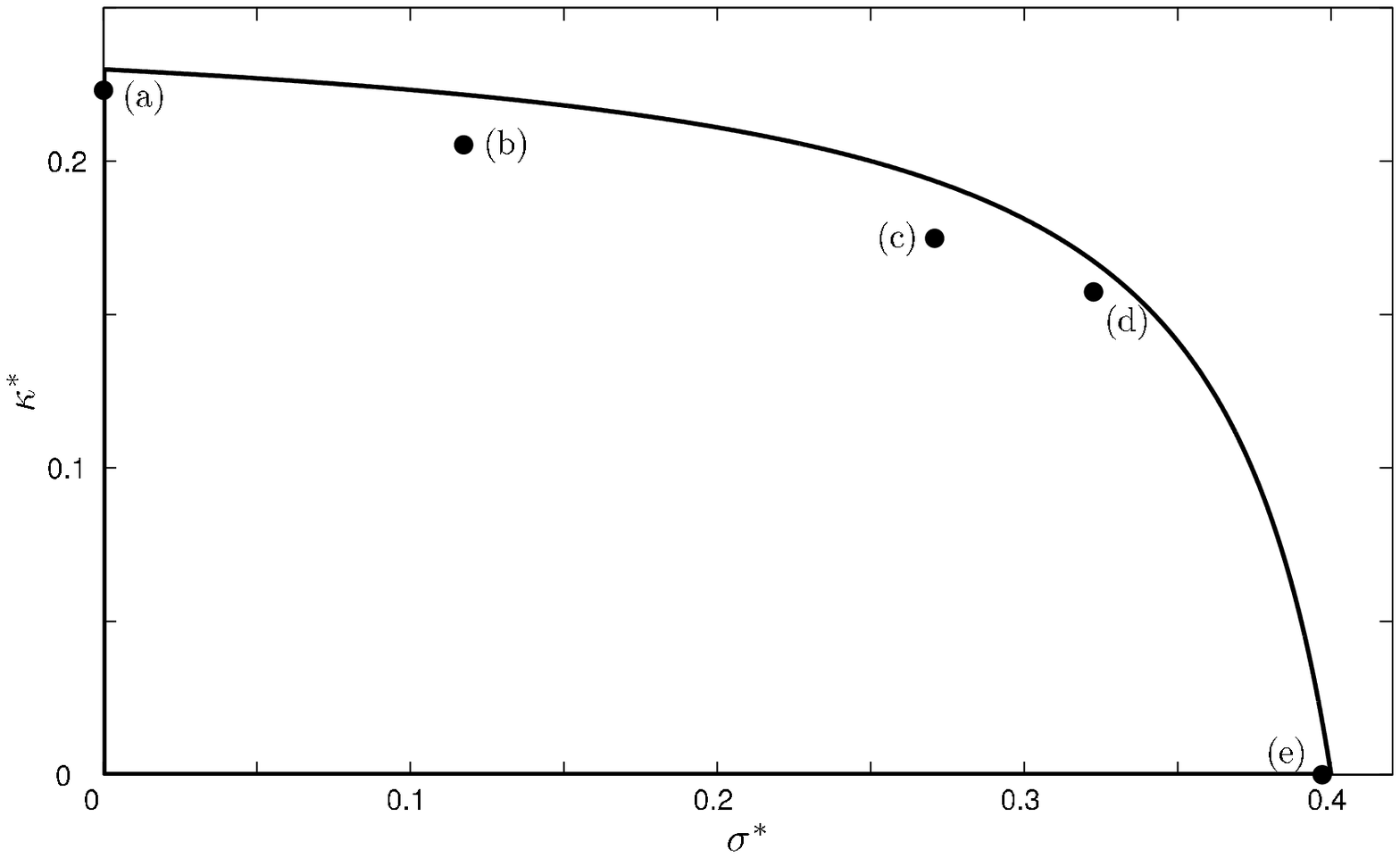}\\
\end{center}
  \caption{Effective properties for the optimized structures with
$V_1=\frac{1}{2}$ given in Fig.~\ref{fig:50_50}.  The curve indicates the
cross-property bounds for the problem.}
  \label{fig:50_50_bounds}
\end{figure}

Optimized unit cells for different $(\ok,\os)$ pairs are shown in
Fig.~\ref{fig:50_50}.  The effective properties for these optimized
microstructures are given in Table~\ref{table:50_50}.
Fig.~\ref{fig:50_50_bounds} shows the effective bulk modulus and conductivity
for the optimized structures along with the cross--property bounds for the
correct volume fractions and phase properties.  The optimized structures are
very close to the cross--property bounds and structures at several points along
the bounds were readily obtained by changing the coefficients $\ok$ and $\os$ in
the objective function.  We note that Fig.~\ref{fig:50_50}(a) resembles the
isotropic, maximum bulk modulus structure presented by \citet{sigmundET00}.
 
\begin{table}[!htb]
  \caption{Effective properties of the five optimized structures with
$V_1=\frac{1}{2}$ shown in Fig.~\ref{fig:50_50}.  }
  \label{table:50_50}
\begin{tabular*}{\columnwidth}{@{\extracolsep{\fill}}c c c c c c c c c}
\hline
Case & $(\ok,\os)$ & $J$ & $\sigma^*$ & $\kappa^*$ & $V_1$ & $\aniso$ & $E^*$ &
$g$ \\ 
\hline
(a) & $(1,0) $ & -0.2231 & 0.0000 & 0.2231 & 0.5000 & 0.0003 & 0.3039 & 0\\ 
\hline
(b) & $(1,\frac{1}{10}) $ & -0.2170 & 0.1174 & 0.2053 & 0.4999 & 0.0014 & 0.2857 & 7 \\ 
\hline
(c) & $(1,\frac{1}{6}) $ & -0.2201 & 0.2709 & 0.1749 & 0.4999 & 0.0013 & 0.2547 & 10\\ 
\hline
(d) & $(1,\frac{1}{2}) $ & -0.3187 & 0.3226 & 0.1574 & 0.5003 & 0.0029 & 0.2379 & 10\\ 
\hline
(e) & $(0,1) $ & -0.3971 & 0.3971 & 0.0000 & 0.4996 & 0.0000 & 0.0000 & 0 \\ 
\hline
\end{tabular*}
\emph{Note:} $g$ is the genus per unit cell, see Section \ref{sec:genus}.
\end{table}

\subsection{30\% stiff volume fraction}

\begin{figure}[!htb]
\begin{center}
\includegraphics[width=3.1cm]{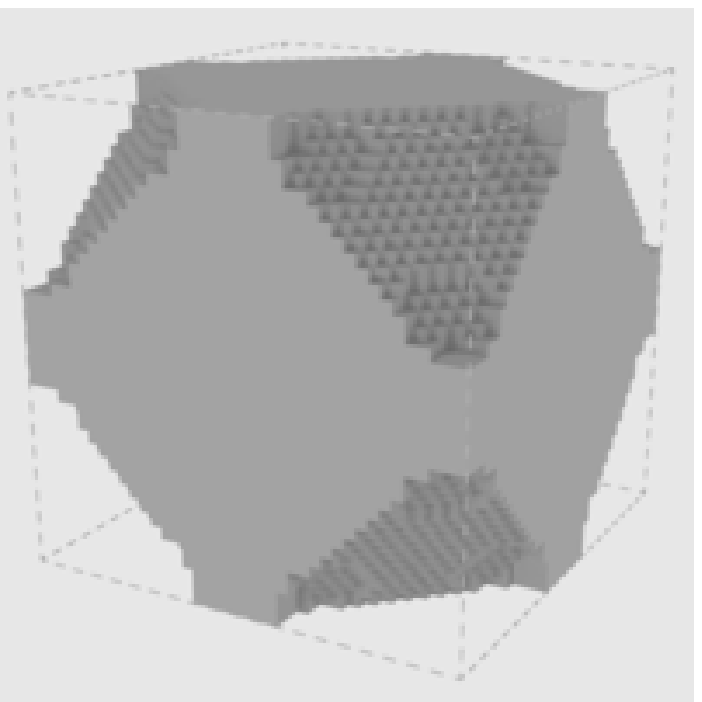}~
\includegraphics[width=3.1cm]{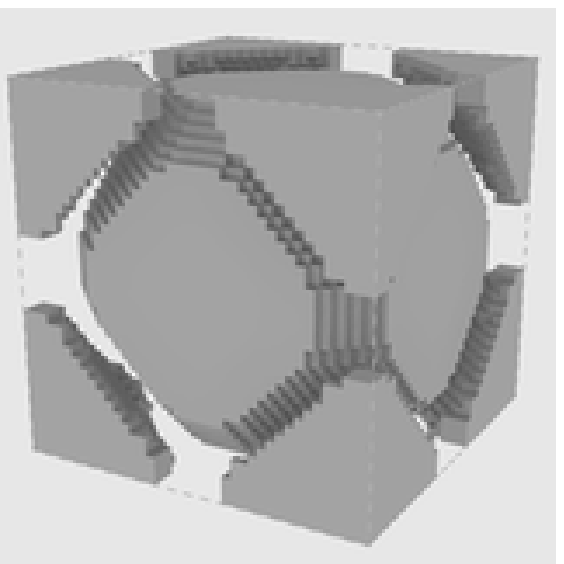}~\hspace{0.5cm}~
\includegraphics[width=3.1cm]{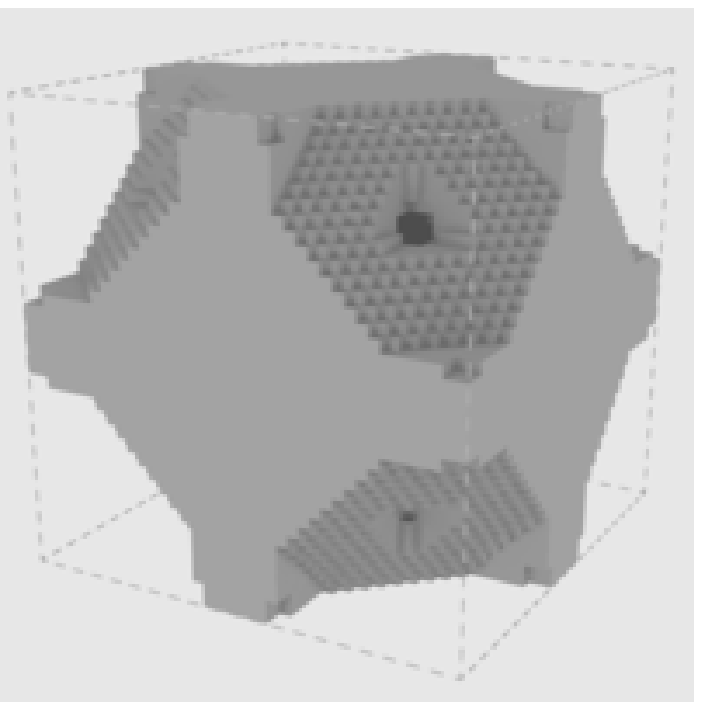}~
\includegraphics[width=3.1cm]{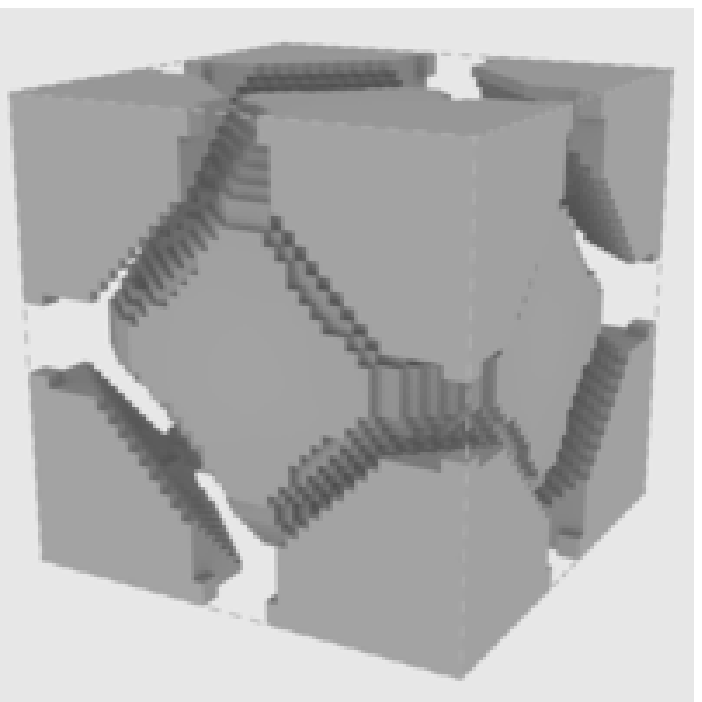}\\
(a) \hspace{6.5cm} (b) \\
\includegraphics[width=3.1cm]{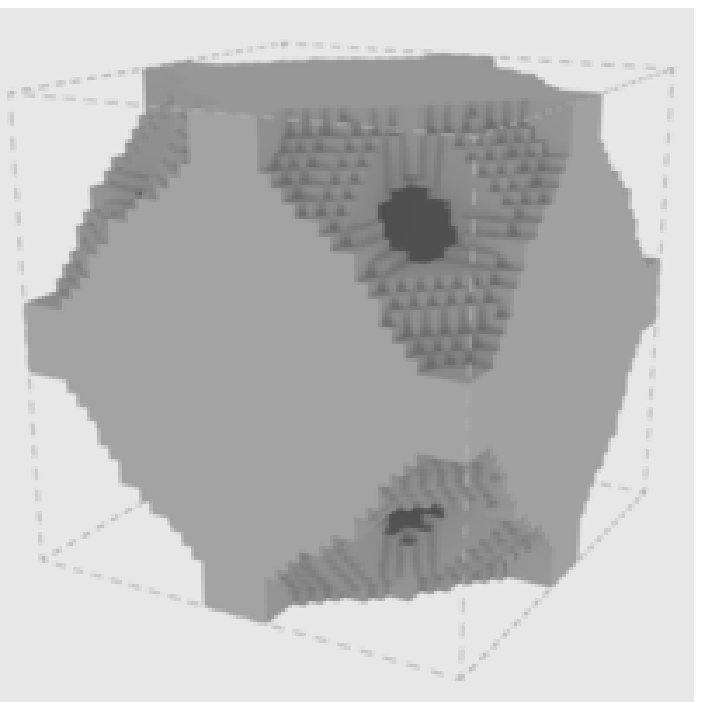}~
\includegraphics[width=3.1cm]{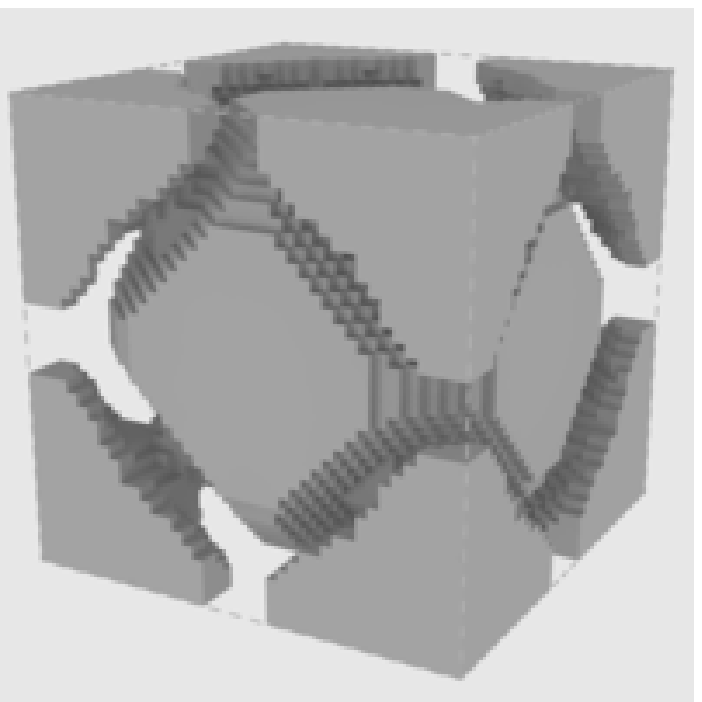}~\hspace{0.5cm}~
\includegraphics[width=3.1cm]{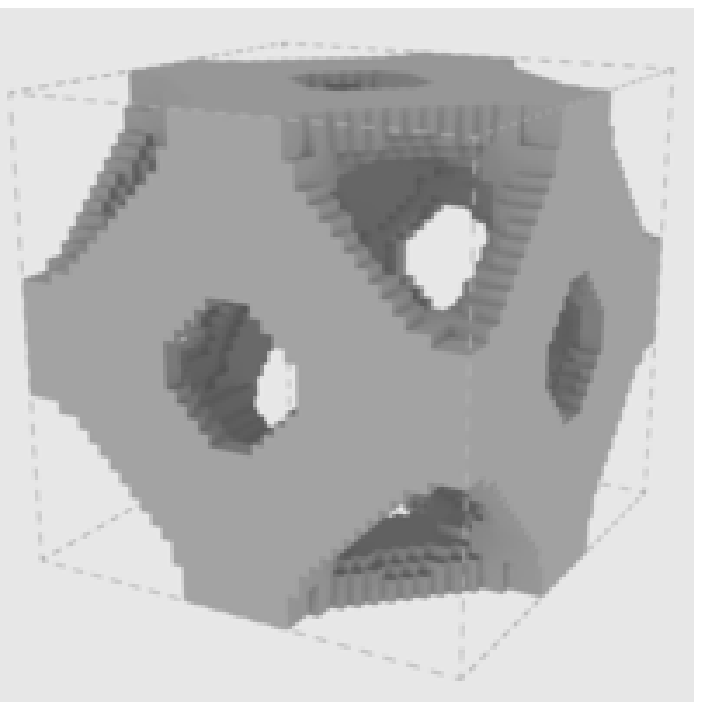}~
\includegraphics[width=3.1cm]{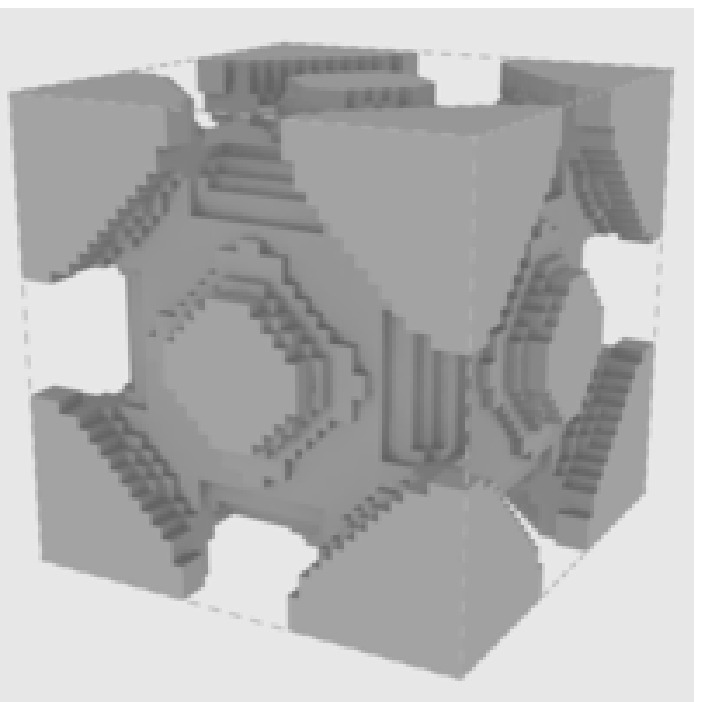}\\
(c) \hspace{6.5cm} (d) \\
\includegraphics[width=3.1cm]{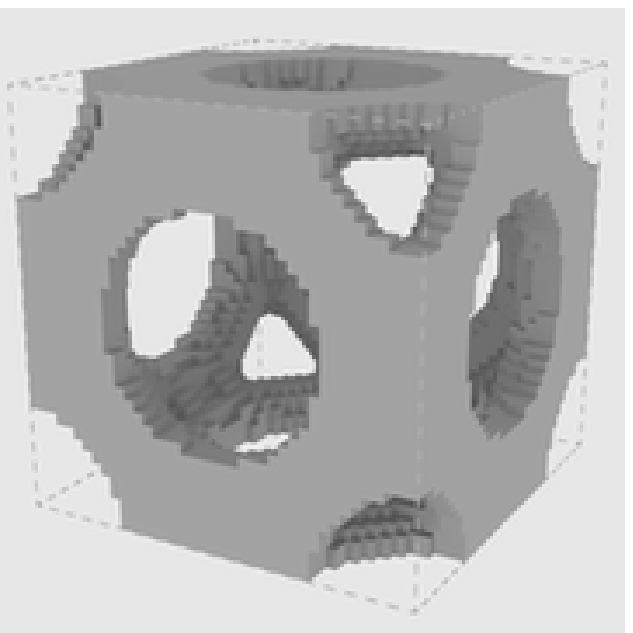}~
\includegraphics[width=3.1cm]{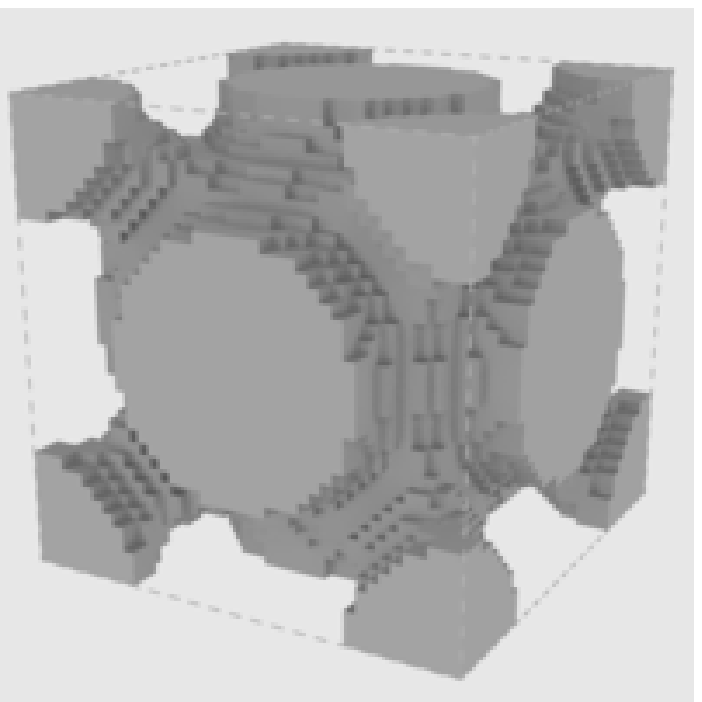}~\hspace{0.5cm}~
\includegraphics[width=3.1cm]{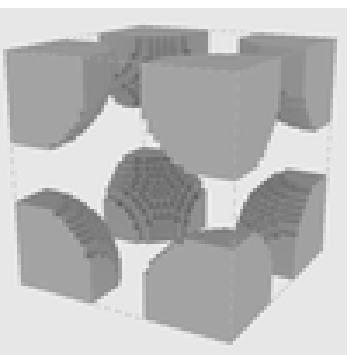}~
\includegraphics[width=3.1cm]{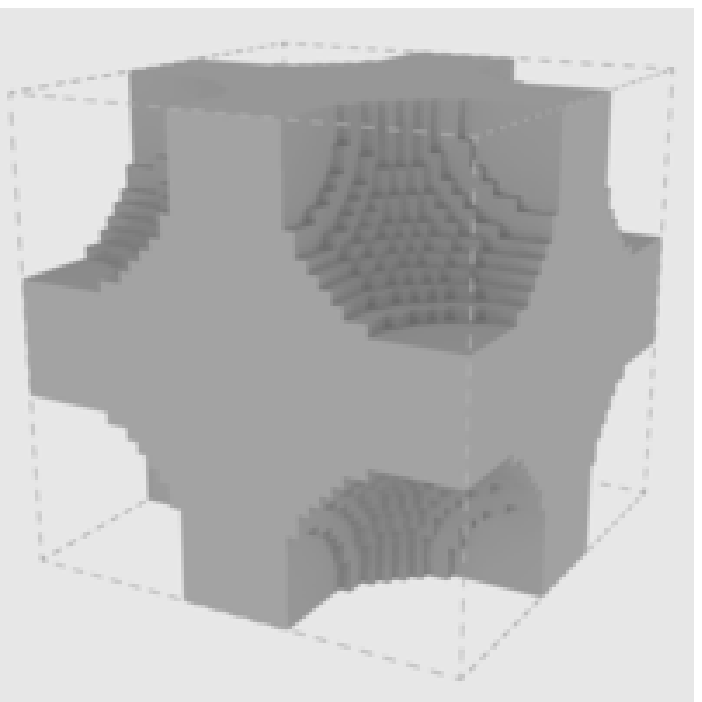}\\
(e) \hspace{6.5cm} (f)
\end{center}
  \caption{Optimized unit cells with $V_1=\frac{3}{10}$ and different weighting
schemes in the objective function: (a) $(\ok,\os)=(1,0)$, (b)
$(\ok,\os)=(1,\frac{1}{50})$, (c) $(\ok,\os)=(1,\frac{1}{20})$, (d)
$(\ok,\os)=(1,\frac{1}{10})$, (e) $(\ok,\os)=(1,\frac{1}{2})$, (f)
$(\ok,\os)=(0,1)$.  In each case the left and right picture show the stiff phase
and conductive phase respectively.}
  \label{fig:30_70}
\end{figure}

\begin{figure}[!htb]
\begin{center}
\begin{minipage}{\textwidth}
\hspace{-1cm}
\includegraphics[width=1.1\textwidth]{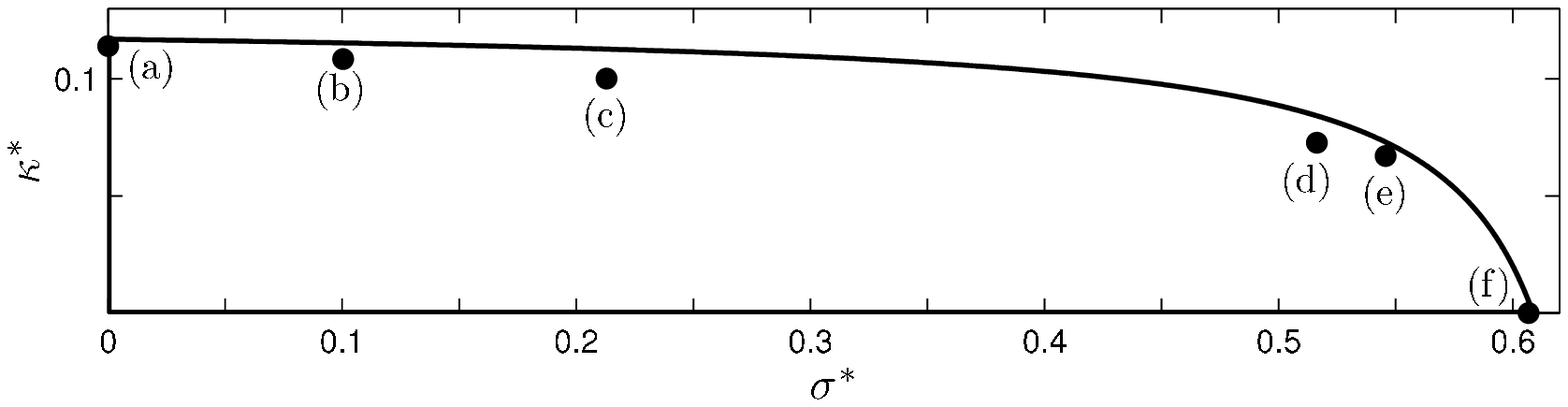}\\
\end{minipage}
\end{center}
  \caption{Effective properties for the optimized structures with
$V_1=\frac{3}{10}$ given in Fig.~\ref{fig:30_70}.  The curve indicates the
cross-property bounds for the problem.}
  \label{fig:30_70_bounds}
\end{figure}

To further explore the capabilities of the level--set method for microstructure
design and the near--optimal structures for the stiffness--conductivity problem,
a study similar to the above was performed with a required stiff phase volume
fraction of $V_1=\frac{3}{10}$.  Optimized structures for $V_1=\frac{3}{10}$ and
different coefficients in the objective function are presented in
Fig.~\ref{fig:30_70}.  The effective properties for these structures are given
in Table~\ref{table:30_70}  and are summarized alongside the correct
cross--property bounds in Fig.~\ref{fig:30_70_bounds}.

\begin{table}[!htb]
  \caption{Effective properties of the optimized structures with
$V_1=\frac{3}{10}$ shown in Fig.~\ref{fig:30_70}.  }
  \label{table:30_70}
\begin{tabular*}{\columnwidth}{@{\extracolsep{\fill}}c c c c c c c c c c}
\hline
Case & $(\ok,\os)$ & $J$ & $\sigma^*$ & $\kappa^*$ & $V_1$ & $\aniso$ & $E^*$ & $g$ \\
\hline
(a) & $(1,0)$ & -0.1140 & 0.0000 & 0.1140 & 0.3000 & 0.0022 & 0.1459 & 0\\
\hline
(b) & $(1,\frac{1}{50})$ & -0.1105 & 0.1004 & 0.1085 & 0.3000 & 0.0004 & 0.1383 & 7 \\
\hline
(c) & $(1,\frac{1}{20})$ & -0.1108 & 0.2129 & 0.1001 & 0.3004 & 0.0009 & 0.1309 & 7\\
\hline
(d) & $(1,\frac{1}{10})$ & -0.1243 & 0.5163 & 0.0727 & 0.3000 & 0.0039 & 0.1006 & 10\\
\hline
(e) & $(1,\frac{1}{2})$ & -0.3399 & 0.5457 & 0.0671 & 0.2996 & 0.0011 & 0.0987 & 10\\
\hline
(f) & $(0,1)$ & -0.6069 & 0.6069 & 0.0000 & 0.2998 & 0.0000 & 0.0000 & 0\\
\hline
\end{tabular*}
\end{table}

\subsection{70\% stiff volume fraction}

\begin{figure}[!htb]
\begin{center}
\includegraphics[width=3.1cm]{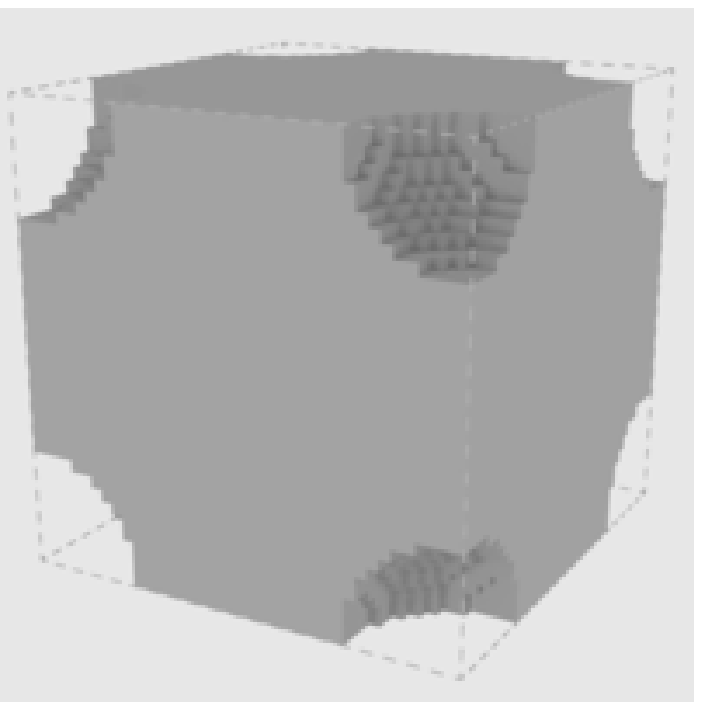}~
\includegraphics[width=3.1cm]{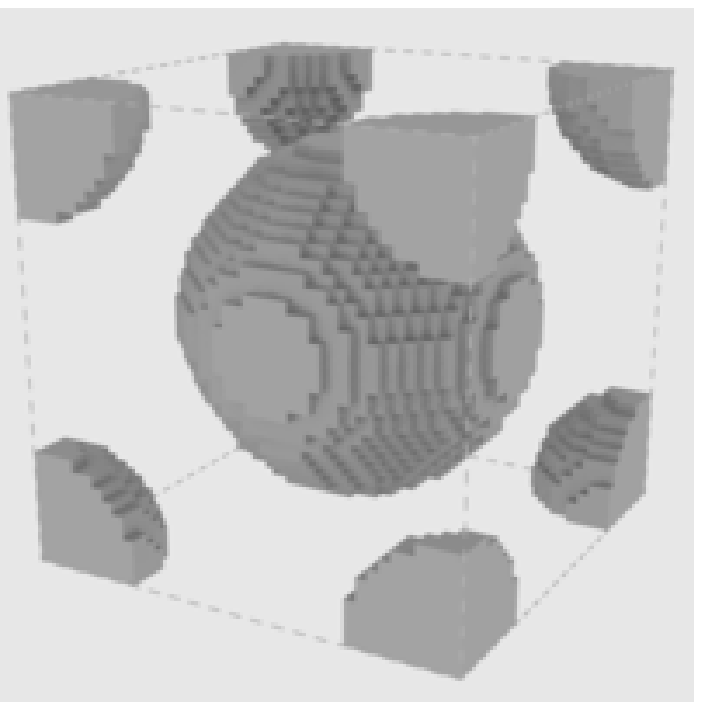}~\hspace{0.5cm}~
\includegraphics[width=3.1cm]{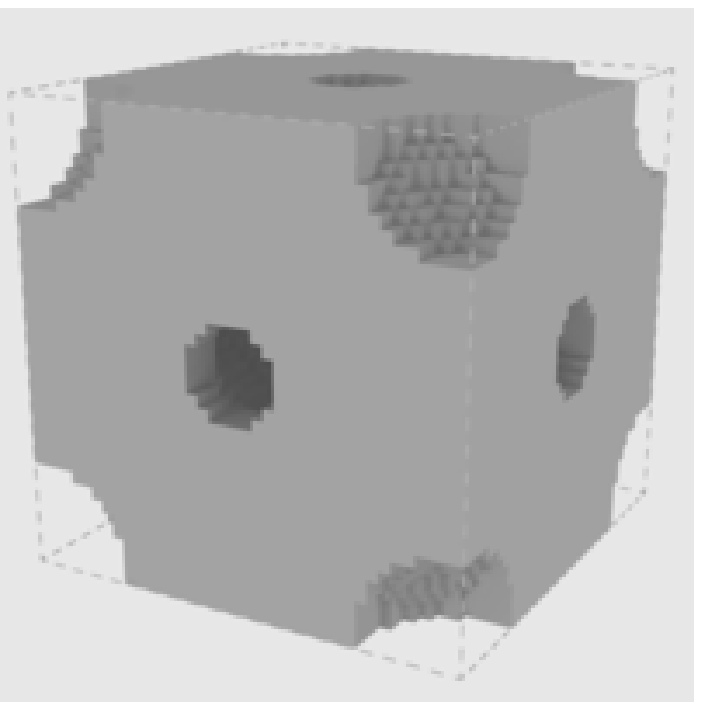}~
\includegraphics[width=3.1cm]{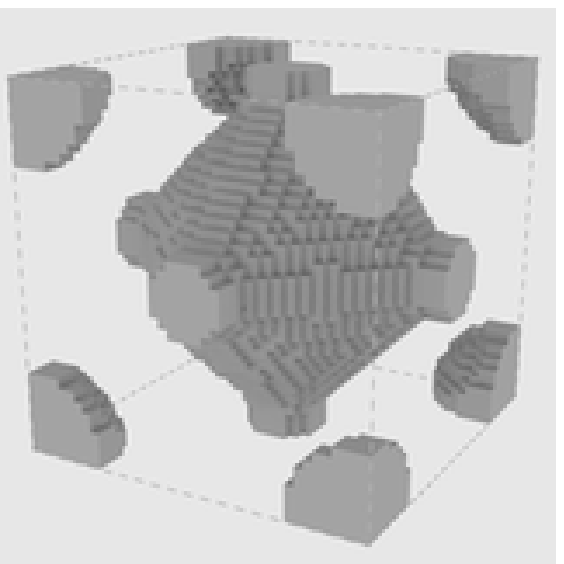}\\
(a) \hspace{6.5cm} (b) \\
\includegraphics[width=3.1cm]{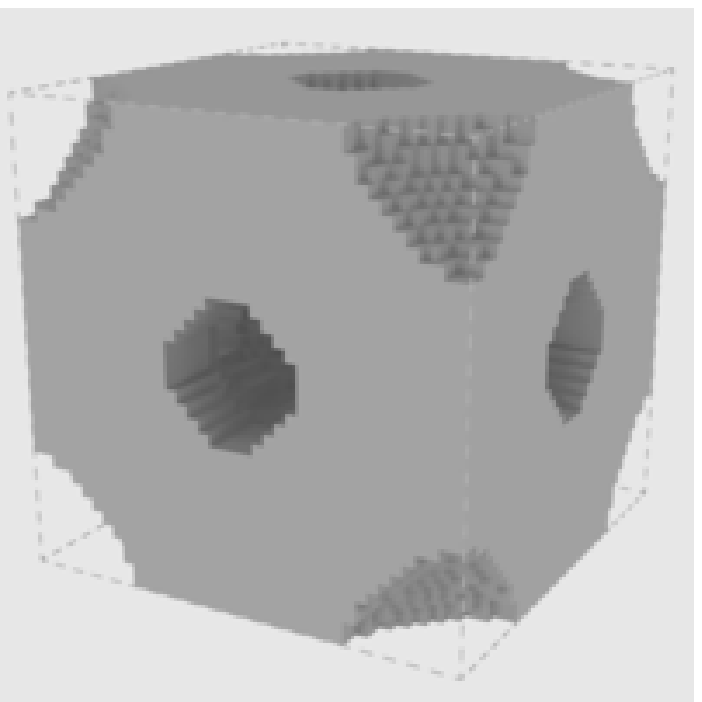}~
\includegraphics[width=3.1cm]{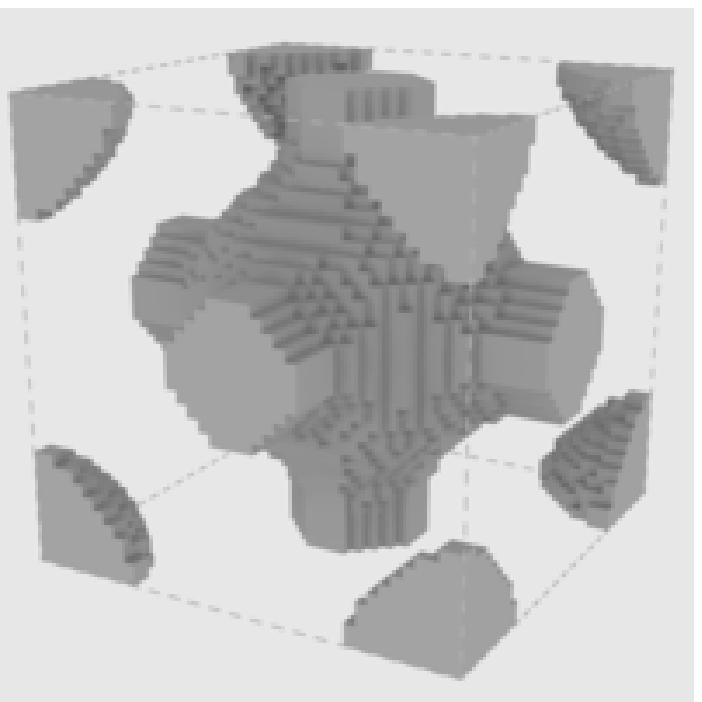}~\hspace{0.5cm}~
\includegraphics[width=3.1cm]{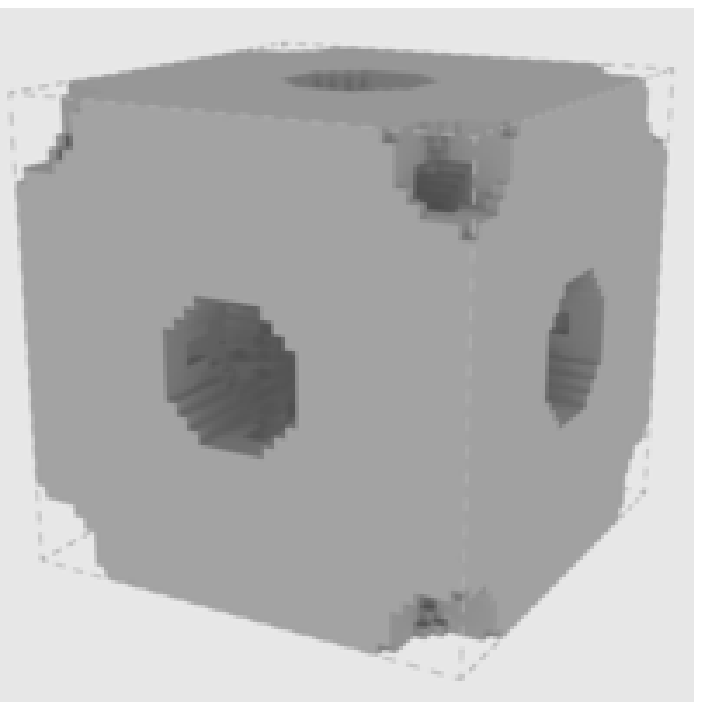}~
\includegraphics[width=3.1cm]{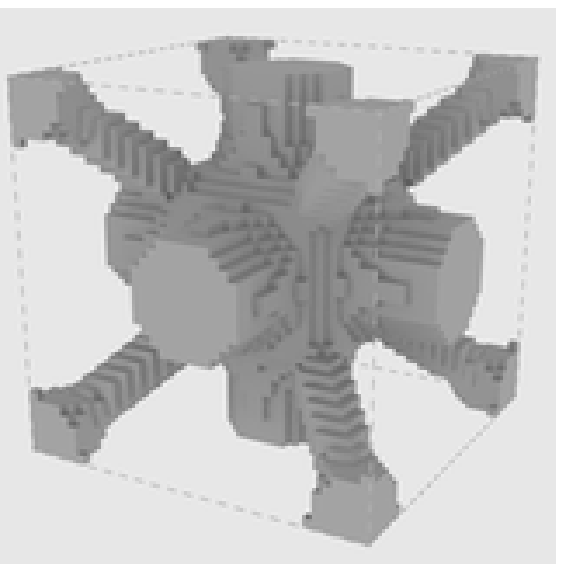}\\
(c) \hspace{6.5cm} (d) 
\end{center}
  \caption{Optimized unit cells with $V_1=\frac{7}{10}$ and different weighting
schemes in the objective function: (a) $(\ok,\os)=(1,0)$, (b)
$(\ok,\os)=(1,\frac{1}{3})$, (c) $(\ok,\os)=(1,\frac{1}{2})$, (d)
$(\ok,\os)=(1,1)$.  In each case the left and right picture show the stiff phase
and conductive phase respectively.}
  \label{fig:70_30}
\end{figure}

\begin{figure}[!htb]
\begin{center}
\includegraphics[width=0.4\textwidth]{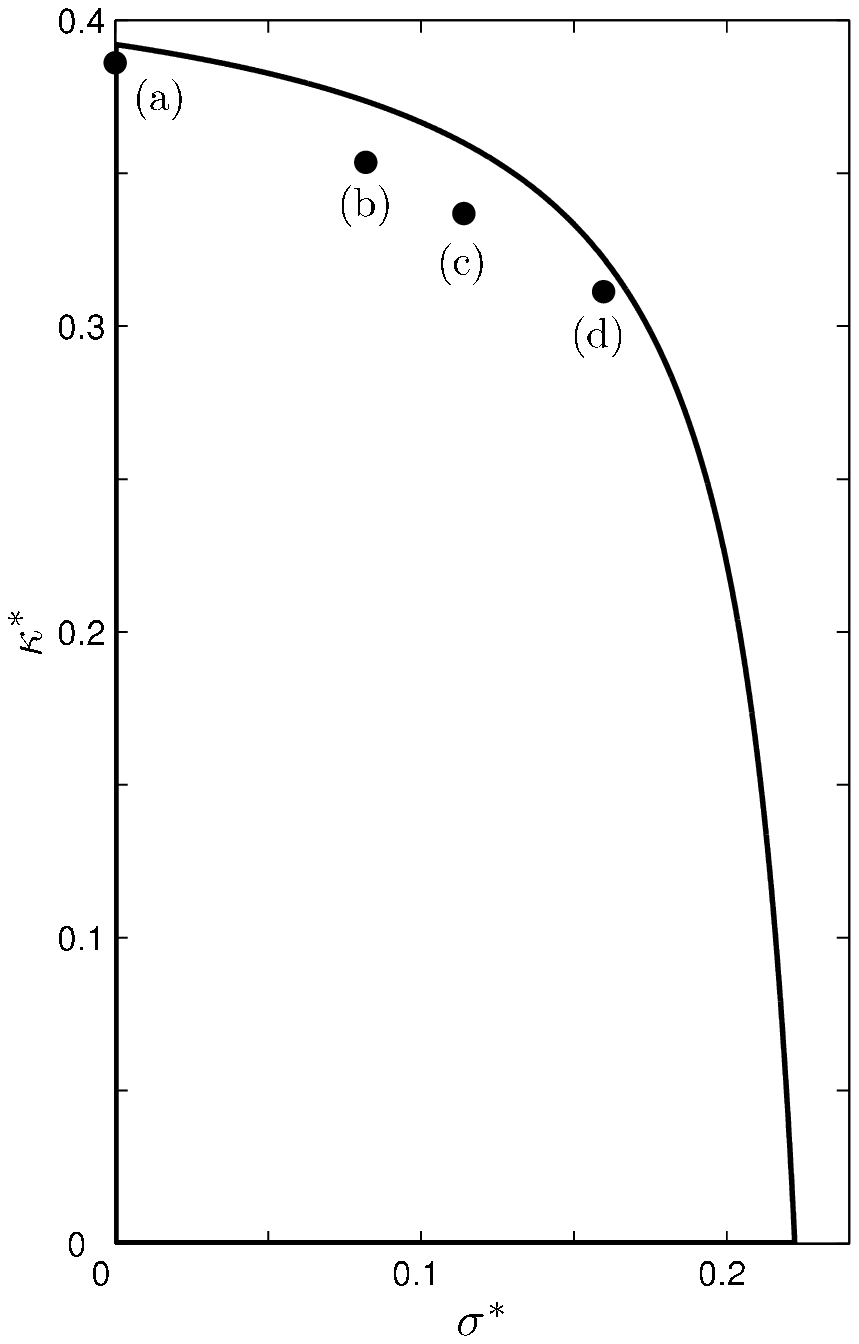}\\
\end{center}
  \caption{Effective properties for optimized structures with $V_1=\frac{7}{10}$
given in Fig.~\ref{fig:70_30}.  The curve indicates the cross-property bounds
for the problem.}
  \label{fig:70_30_bounds}
\end{figure}

This section presents optimized structures with a required stiff phase volume
fraction of $V_1=\frac{7}{10}$.  Optimized structures are displayed in
Fig.~\ref{fig:70_30}.  Effective properties for the optimized structures are
presented in Table~\ref{table:70_30} and summarized alongside the
cross--property bounds in Fig.~\ref{fig:70_30_bounds}.

\begin{table}[!htb]
  \caption{Effective properties of the optimized structures with
$V_1=\frac{7}{10}$ shown in Fig.~\ref{fig:70_30}.  }
  \label{table:70_30}
\begin{tabular*}{\columnwidth}{@{\extracolsep{\fill}}c c c c c c c c c c}
\hline
Case & $(\ok,\os)$ & $J$ & $\sigma^*$ & $\kappa^*$ & $V_1$ & $\aniso$ & $E^*$ & $g$ \\
\hline
(a) & $(1,0)$ & -0.3861 & 0.0000 & 0.3861 & 0.6998 & 0.0001 & 0.5234 & 0\\
\hline
(b) & $(1,\frac{1}{3})$ & -0.3808 & 0.0819 & 0.3536 & 0.6999 & 0.0006 & 0.4940 & 3\\
\hline
(c) & $(1,\frac{1}{2})$ & -0.3938 & 0.1140 & 0.3368 & 0.7003 & 0.0004 & 0.4762 & 3\\
\hline
(d) & $(1,1)$ & -0.4710 & 0.1597 & 0.3113 & 0.7001 & 0.0012 & 0.4547 & 10\\
\hline
\end{tabular*}
\end{table}

\section{Discussion}
\label{sec:genus}
\label{discussion}

The optimized structures with $V_1=\frac{1}{3}$  display the same set of
topologies as the optimized structures with $V_1=\frac{1}{2}$.  The optimized
structures for both cases have effective properties very close to the
cross--property bounds.  

As in Fig.~\ref{fig:egoptim}, it is frequently the conductive phase which
extends out to make new topologies in the optimization process.  Accordingly, it
was more difficult to obtain optimized structures close to the cross--property
bounds for the $V_1=\frac{7}{10}$ case.  We see familiar topologies from the
other volume fraction cases for structures (a) and (d) in Fig.~\ref{fig:70_30}.
However, along the low conductivity section of the curve the optimized
structures ((b) and (c) in Fig.~\ref{fig:70_30}) have a conductive phase which
is in two disconnected parts.  One part of the conductive phase provides
effective conductivity whereas the other is present to satisfy the isotropy
constraint.  

For all three volume fraction cases we have been unable to obtain structures in
the high--conductivity part of the cross--property bounds curve which have
nonzero stiffness.  For example, see the gap between structures (d) and (e) in
Fig.~\ref{fig:50_50_bounds}.  We believe this problem stems from the isotropy
constraint: for a small effective stiffness the algorithm simply disconnects the
stiff phase to make it trivially isotropic.  Aiming for this section of the
curve $\os>\ok$ in the objective function, thus zero stiffness does not affect
the objective function significantly.  Structures in this region may be found by
using an objective function which includes a penalty for disconnecting each of
the phases, e.g., $J=\frac{\omega_\kappa}{\kappa^*} +
\frac{\omega_\sigma}{\sigma^*}$.  Alternatively, significant experimentation
with $\al_{min}^2$ may give structures with the desired properties.
  
For the $V_1=\frac{7}{10}$ case we were also unable to obtain a structure close
to $(\sigma_{HS},0)$.  The geometry expected for this structure given those
obtained for the other volume fraction cases appears impossible at
$V_1=\frac{7}{10}$.  Swapping the  phases of structure (a) from
Fig.~\ref{fig:30_70}, which has $V_1=\frac{3}{10}$ and properties close to
$(0,\kappa_{HS})$, would give a structure very close to $(\sigma_{HS},0)$ on the
$V_1=\frac{7}{10}$ cross--property bounds.  However, it is the isotropy
requirement for the elastic phase which drives the algorithm to the
body--centered arrangement of Fig.~\ref{fig:30_70}(a).  Given that the algorithm
moves towards disconnecting the stiff phase, the optimization is not driven
towards the correct structure.  Instead it tends towards the simple cubic
structure seen at other volume fractions, and when it is unable to disconnect
the stiff phase cannot then move towards the body--centered geometry.  

\begin{figure}
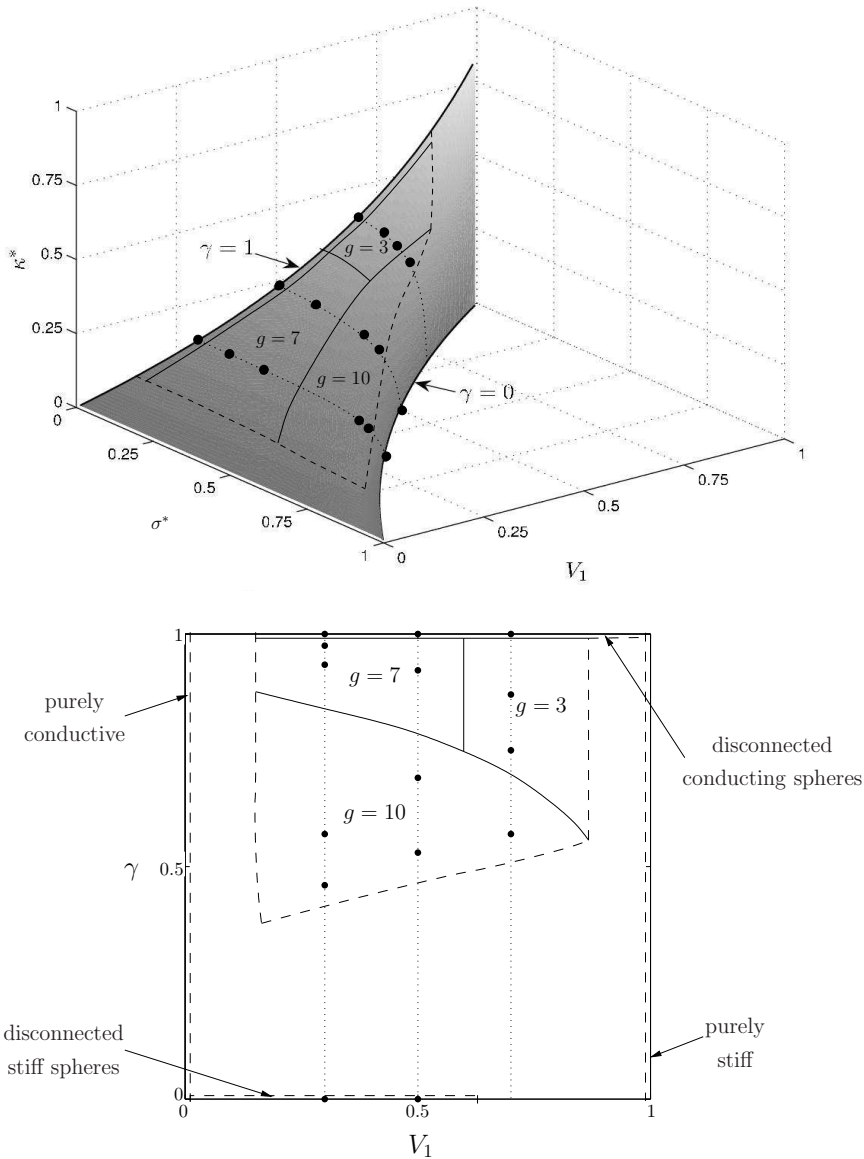

\hspace{1.5cm}
\begin{minipage}{10cm}
\scalebox{0.6}{ \input{3d_phase_diagram.pstex_t}}\\
\scalebox{0.7}{ \input{phase_diagram_flipped.pstex_t}}
\end{minipage}
\caption{The effective properties of optimal stiff and conducting
microstructures for all volume fractions (above) and a parameterization of this
surface using the stiff phase volume fraction $V_1$ and the parameter $\gamma$
from Eq.~\ref{eq:hyp} (below). In each case the black dots correspond to
particular microstructures inferred from the optimized structures we have
presented.  The lower ``phase--diagram'' is divided into areas for which we
hypothesize the existence of optimal microstructures with a particular topology.
The three vertical dotted lines represent the three stiff phase volume fractions
considered in this paper.  Solid lines represent boundaries between regions, but
should be considered as approximate. Dashed lines represent the extent of what
we can infer from the optimized structures presented.  Corresponding regions can
be seen on the surface in the upper diagram.    } \label{fig:phase_diagram}
\end{figure}

In Fig.~\ref{fig:phase_diagram} we hypothesize the existence of optimal
microstructures with the topologies of the optimized microstructures we have
presented.  The microstructures are grouped based on their topology using the
genus $g$ which represents the number of \emph{handles} of an object
\citep{hyde89}.   



\section{Concluding remarks}
\label{conclusion}

This paper presents a method for material design using the level--set method of
topology optimization.  The method is utilized to design isotropic periodic
composite materials in three dimensions which are maximally stiff and conducting
from two ill--ordered, infinite contrast phases.  The optimized microstructures
presented here have properties very close to the relevant conductivity bulk
modulus cross--property bounds, proving the capabilities of the level--set
method for microstructure design.  

Until now, the only known optimal single--scale microstructures for this problem
are cubic symmetric and therefore have weak directions.  An important component
of our method is the ability to improve upon this by imposing an isotropy
constraint.  The isotropy of the optimized microstructures makes them attractive
for engineering applications: provided the shear modulus of the microstructures
is high, the composites will have a high Young's modulus in all directions.  The
isotropy requirement means that almost all of our optimized microstructures have
not been presented previously.  Our method can readily be applied to other
constraints. 
 
For various volume fractions we hypothesize the existence of optimal
single--scale microstructures with the topologies of those presented.  We have
not been able to find near--optimal structures for the high--conductivity
nonzero stiffness section of the cross--property diagram.  This is a result of
the the isotropy constraint; instead of constraining the microstructures to be
isotropic while being only weakly stiff the algorithm disconnects the stiff
phase to make it trivially isotropic.   

The freedom to choose the initial unit cell has not been discussed.  Different
initial cells can give different prescribed symmetries and preliminary
investigations demonstrate that different initial cells will almost certainly
result in different optimized microstructures close to the cross--property
bounds.  Thus there are possibly other microstructures with similar properties
to those presented; this would provide the designer with freedom to choose the
microstructure to suit their purposes.  However, given the closeness of our
optimized structures to the cross--property bounds, other microstructures found
could not outperform them by much more than a few percent.  

\section*{Acknowledgements}
The authors thank Dr.~James K. Guest for reading the manuscript and providing
useful comments.  This work was supported by a grant from the Australian
Research Council through the Discovery Grant scheme, and an Australian
Postgraduate Award.  Computational resources used in this work were provided by
the Queensland Cyber Infrastructure Foundation.

\bibliographystyle{elsart-harv}
\bibliography{refs}

\end{document}